\documentclass[11pt]{article}
\usepackage{a4wide}
\usepackage[normalem]{ulem}
\usepackage{graphicx}
\usepackage{amsmath,amssymb}
\usepackage{color}
\usepackage{slashed}
\usepackage[bookmarks=true,bookmarksnumbered=true,setpagesize=false]{hyperref}

\usepackage{comment}

\newcommand{\tx}{\text}

\newcommand{\nn}{\nonumber\\}
\newcommand{\h}{\hspace}
\newcommand{\be}{\begin{equation}}
\newcommand{\e}{\end{equation}}
\newcommand{\aln}[1]{\begin{align}#1\end{align}}

\newcommand{\ov}{\over}

\begin{document}
\title{
\vspace{-2cm}
\vbox{
\baselineskip 14pt
\hfill \hbox{\normalsize OU-HET-1087
}} 
\vskip 1cm
\bf \Large Minimal scenario of Criticality for Electroweak scale, Neutrino Masses, Dark Matter, and Inflation
\vskip 0.5cm
}
\author{
Yuta Hamada,\thanks{E-mail: \tt yhamada@fas.harvard.edu}\,
Hikaru~Kawai,\thanks{E-mail: \tt hikarukawai@phys.ntu.edu.tw}\, 
Kiyoharu Kawana,\thanks{E-mail: \tt kawana@snu.ac.kr}\,
Kin-ya Oda,\thanks{E-mail: \tt odakin@lab.twcu.ac.jp}\, and
Kei Yagyu\thanks{E-mail: \tt yagyu@het.phys.sci.osaka-u.ac.jp}
\bigskip\\
\normalsize
\it 
 $^*$  Department of Physics, Harvard University, Cambridge, MA 02138 USA,\\
 \normalsize 
 \it
$^{\dagger}$ Department of Physics and Center for Theoretical Physics,
 \\
 \normalsize
\it 
  National Taiwan University, Taipei, Taiwan 106,\\
 \normalsize 
 \it 
  Physics Division, National Center for Theoretical Sciences, Taipei 10617, Taiwan 
\\ 
\normalsize 
\it  
$^{\ddag}$  Center for Theoretical Physics, Department of Physics and Astronomy,\\
 \normalsize
\it  Seoul National University, Seoul 08826, Korea,
\\
\normalsize
\it
$^{\S}$ Department of Mathematics, Tokyo Woman’s Christian University, Tokyo 167-8585, Japan,
\\
\normalsize
\it  $^{\P}$ Department of Physics, Osaka University, Osaka 560-0043, Japan
\smallskip
}

\date{\today}

\maketitle

\begin{abstract}\noindent
We propose a minimal model that can explain the electroweak scale, neutrino masses, Dark Matter (DM), and successful inflation all at once based on the multicritical-point principle (MPP).  
The model has two singlet scalar fields that realize an analogue of the Coleman-Weinberg mechanism, in addition to the Standard Model with heavy Majorana right-handed neutrinos. 
By assuming a $Z_2 $ symmetry, one of the scalars becomes a DM candidate whose property is almost the same as the minimal Higgs-portal scalar DM. 
In this model, the MPP can naturally realize a saddle point in the Higgs potential at high energy scales.  
By the renormalization-group analysis, we study the critical Higgs inflation with non-minimal coupling $\xi |H|^2 R$ that utilizes the saddle point of the Higgs potential.
We find that it is possible to realize successful inflation even for $\xi=25$ and that the heaviest right-handed neutrino is predicted to have a mass around $10^{14}$ ${\rm GeV}$ to meet the current cosmological observations. 
Such a small value of $\xi$ can be realized by the Higgs-portal coupling $\lambda_{SH}\simeq 0.32$ and the vacuum expectation value of the additional neutral scalar $\langle\phi\rangle\simeq 2.7$\,TeV, which correspond to the dark matter mass 2.0\,TeV, its spin-independent cross section $1.8\times10^{-9}$\,pb, and the mass of additional neutral scalar 190\,GeV. 
\end{abstract}
\newpage

\section{Introduction}
The observed Higgs mass supports the assumption that the Standard Model (SM) is not much altered up to the Planck scale.
More precisely, the critical value of the top-quark pole mass is about $m_{t,\tx{critical}}^{\rm pole}\simeq 171.4$ GeV~\cite{Hamada:2014wna} for the theoretical border between stability and instability (or metastability) of the effective Higgs potential for the observed Higgs mass $\simeq$125\,GeV; see also Refs.~\cite{Degrassi:2012ry,Buttazzo:2013uya,Bednyakov:2015sca}.\footnote{
With the current central value $m_H=125.1\pm0.1$\,GeV~\cite{PDG2020}, the critical top mass becomes $m_{t,\tx{critical}}^{\rm pole}\simeq171.2$\,GeV~\cite{Bednyakov:2015sca}. 
}
This critical value of the top pole mass is consistent at the 1.4\,$\sigma$ level with the latest combination of the experimental results $m_t^{\rm pole}=172.4\pm 0.7$ GeV~\cite{PDG2020}. 
Surprisingly, the degenerate minimum of the Higgs potential at the critical top mass coincides with the Planck scale, and such a behavior of the potential has a lot of implications to high energy physics and cosmology.  
This interesting behavior of the Higgs potential can be understood by the multicritical-point principle (MPP)~\cite{Froggatt:1995rt,Froggatt:2001pa,Nielsen:2012pu,Kawai:2011rj,Kawai:2011qb,Kawai:2013wwa,Hamada:2014ofa,Hamada:2014xra,Hamada:2015dja,Hamada:2015ria,Kannike:2020qtw} (see also Refs.~\cite{Bennett:1988xi,Bennett:1993pj}) that ``coupling constants that are relevant at low energies are tuned to a multicritical point around which the vacuum structure drastically changes when they are varied.''
We note that existence of a saddle-point around the Planck scale, rather than a degenerate vacua, is another possible form of multicriticality. This fact is used in the critical Higgs inflation~\cite{Hamada:2014iga,Bezrukov:2014bra,Hamada:2014wna,Hamada:2014xka,Hamada:2017sga} explained below.

Besides such interesting behavior of the Higgs potential, there are many mysteries and problems in particle physics and cosmology.   
For example, we have not yet understood the origin of electroweak (EW) scale $v=246$ GeV, which is hugely small compared to the Planck or string  scale $10^{18}$ GeV at which people believe that there must exist an unified theory which includes quantum gravity.  
Further, the Majorana-mass scale for the right-handed neutrinos is  unknown in the SM with the seesaw mechanism~\cite{Minkowski:1977sc,Yanagida:1979as,GellMann:1980vs,Glashow:1979nm,Mohapatra:1979ia}. 
Moreover, the recent observations in cosmology, including that of the cosmic microwave background (CMB), have established the existence of (cold) dark matter (DM).
It motivates us to consider a new particle whose interactions with the SM particles are relatively weak.            

In addition, the CMB fluctuations may also provide hints for further new physics since they are seeded at high energy scales during inflation. 
Current observation is consistent with single-field inflation models, among which the Higgs inflation provides one of the best fits~\cite{Bezrukov:2007ep,Barvinsky:2008ia,DeSimone:2008ei,Allison:2013uaa}.  
In particular, the critical Higgs inflation is inspired by the possible existence of the saddle point around the Planck scale in the MPP, which helps to flatten the Higgs potential and allows rather small value of the 
non-minimal coupling $\xi={\cal O}(10)$ in $\xi |H|^2R$. (The not-so-large coupling is favorable from unitarity~\cite{Burgess:2009ea,Barbon:2009ya,Burgess:2010zq,Bezrukov:2010jz,Ema:2016dny}.)

In this paper, we consider the most economical model that can simultaneously explain all the above issues: the critical Higgs inflation, neutrino masses, EW scale, and DM.\footnote{Quite recently, 
a similar scenario to simultaneously explain these problems has been studied in Ref.~\cite{Kubo:2020fdd} in the context of the classical scale invariance, in which a fermionic DM candidate is considered.
In our scenario, we consider the MPP and a scalar DM candidate.}
The model consists of two additional real singlet scalar fields and the SM with the right-handed neutrinos.
As a result, we manage to predict the parameters in a well determined narrow region by taking into account the constraints from the DM relic abundance, its direct detection experiments, the CMB fluctuations, and the latest LHC data, while keeping the perturbativity up to the Planck scale. 
We find that the Higgs-portal coupling and the vacuum expectation value of the additional neutral scalar are fixed to be $\lambda_{SH}\simeq 0.32$ and $\langle\phi\rangle\simeq 2.7$\,TeV, resulting in the dark matter mass 2.0\,TeV, its spin-independent cross section $1.8\times10^{-9}$\,pb, and the mass of additional neutral scalar 190\,GeV.

The motive behind this model is as follows.
If $m_t^\tx{pole}>m_{t,\tx{critical}}^\tx{pole}$, the instability of the Higgs potential requires additional positive contributions to the renormalization group (RG) of the Higgs quartic coupling $\lambda_H $.
The simplest possibility is to introduce a real scalar field $S$ that couples to the SM Higgs doublet $H$ via a quartic interaction $\lambda_{SH} |H|^2 S^2$. (This is nothing but the Higgs-portal DM model~\cite{Silveira:1985rk,McDonald:1993ex,Burgess:2000yq,Cline:2013gha}.) 
However, such an extension could yield too large a tensor-to-scalar ratio of the CMB in the (critical) Higgs inflation, because of the raised Higgs potential at high scales; see e.g.~\cite{Hamada:2013mya,Hamada:2017sga}. 
We can resolve this issue by introducing additional superheavy fermions that lowers the Higgs potential at higher scales, while keeping the above mentioned positive contribution from the scalar at lower scales.
Therefore, it is reasonable to assume a high-scale seesaw in which right-handed neutrinos have large Majorana masses $M_R $ and large Yukawa couplings $y_\nu $.
In fact, it is possible to maintain the saddle point of the Higgs potential at high energy scales in the existence of new scalar field(s) when $M_R \sim 10^{14}$ GeV~\cite{Hamada:2017sga}.
This is also one of the interesting predictions of the MPP.  

So far, the MPP has not explained the origin of the EW scale. 
In fact, we may do so as follows. 
It is known that the Coleman-Weinberg (CW) mechanism~\cite{Coleman:1973jx} can naturally explain the hierarchy between the EW and Planck scales through the dimensional transmutation. 
Important assumption behind the CW mechanism is that the renormalized mass-squared parameter vanishes at the origin of the scalar field space. 
This assumption is called the classical scale invariance (CSI)~\cite{Meissner:2006zh,Foot:2007iy,Iso:2009ss,Iso:2009nw,Hur:2011sv,Iso:2012jn,Englert:2013gz,Hashimoto:2013hta,Holthausen:2013ota,Hashimoto:2014ela,Kubo:2014ova,Kubo:2015cna,Jung:2019dog}.\footnote{
In Refs.~\cite{Chankowski:2014fva,Meissner:2018mvq}, the bare Higgs mass too is required to vanish; see also Ref.~\cite{Veltman:1980mj,Hamada:2012bp} for the discussion on bare mass in the SM.
}
On the other hand, such CSI-type potentials can be naturally understood as one of the possible multicriticities in MPP without referring to scale invariance~\cite{Haruna:2019zeu,Hamada:2020wjh}.\footnote{
Throughout this paper, we call a potential that has vanishing first and second derivatives at the origin in the field space, $V'|_{\phi=0}=V''|_{\phi=0}=0$, the CSI-type potential.
}
From this MPP point of view, the simplest realization of the dimensional transmutation is achieved by only two real singlet scalar fields, one of which is $S$ mentioned above~\cite{Haruna:2019zeu,Hamada:2020wjh}.

Although we will focus on a specific model proposed in \cite{Haruna:2019zeu,Hamada:2020wjh}  (namely only the CP 2-2 among various multicritical points in the parameter space) in the following, the analysis of Higgs inflation does not much depend on the details of the model because only the scalar coupling $\lambda_{SH} $ and the neutrino Yukawa $y_\nu $ play important roles to determine the behaviours of the Higgs potential at high energy scales.  
In this sense, the same analysis is easily applicable to similar extensions of the SM.      

The organization of the paper is as follows. 
In Sec.~\ref{model}, we briefly explain the minimal model of dimensional transmutation \cite{Haruna:2019zeu,Hamada:2020wjh} extended with right-handed neutrinos and study the RG.   
In Sec.~\ref{saddle point section}, we study the saddle point of the Higgs potential at high scales.
In Sec.~\ref{inflation}, we discuss the critical Higgs inflation. 
In Sec.~\ref{Prediction on inflationary observables}, we show the method and results for our numerical prediction for the inflationary observables.
Summary and discussion are given in Sec.~\ref{summary}. 
In \ref{RGEs}, we list the two-loop renormalization group equations (RGEs).  
In \ref{Single-field slow-roll inflation}, we summarize basic results for a general single-field slow-roll inflation.
In \ref{ordinary Higgs inflation results}, we summarize basic results for the ordinary (non-critical) Higgs inflation.
In \ref{Expansion around saddle point}, we show analytic results of expansion around the saddle point.

\section{Model}\label{model}
In this section, we introduce the minimal model for the EW scale, neutrino masses, DM and the critical Higgs inflation. 
The model is based on \cite{Haruna:2019zeu,Hamada:2020wjh} whose scalar sector contains two additional real singlet scalars $S$ and $\phi$. 
We also take into account heavy Majorana right-handed neutrinos $\nu_R^i$ with three generations. 
%
In order to make $S$ a DM candidate, we impose the $Z_2 $ symmetry $S\rightarrow -S$, with all the other fields being invariant. 
\begin{figure}[t!]
\begin{center}
\includegraphics[width=7cm]{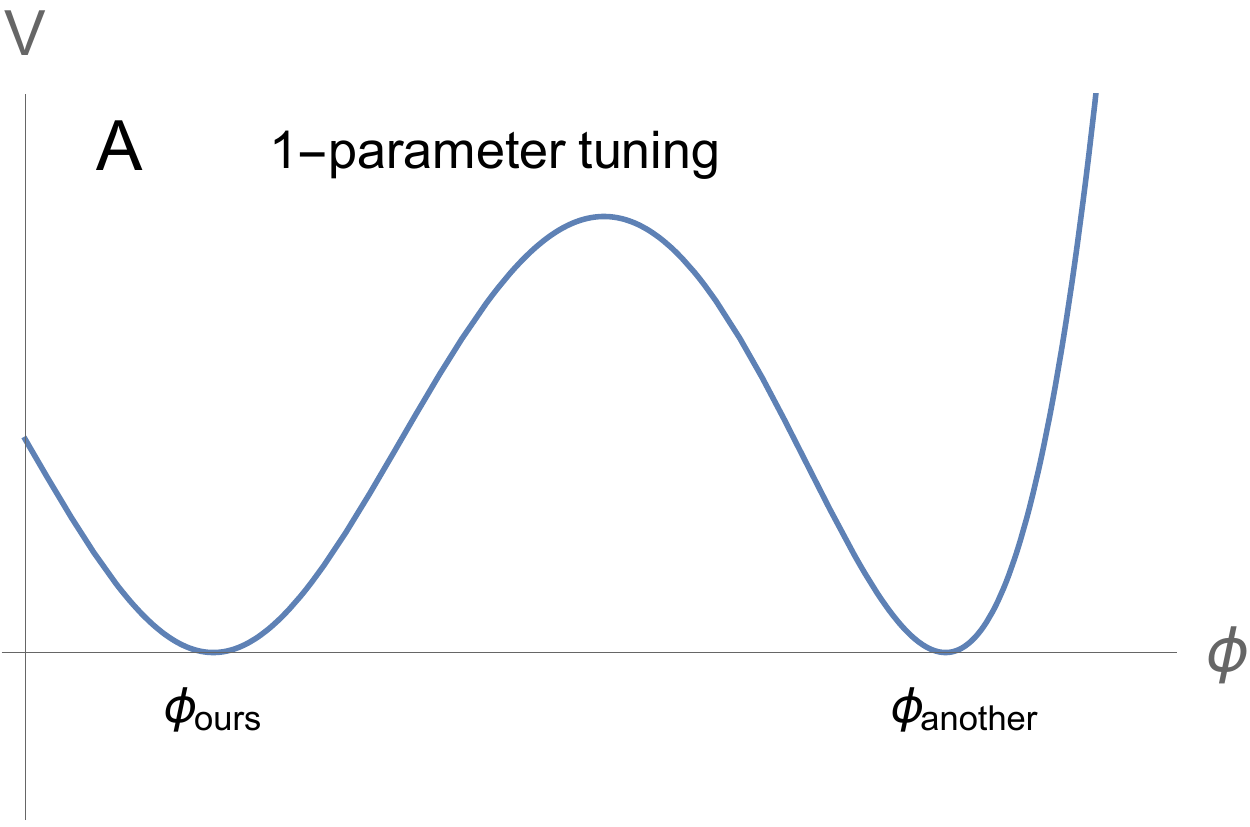}
\includegraphics[width=7cm]{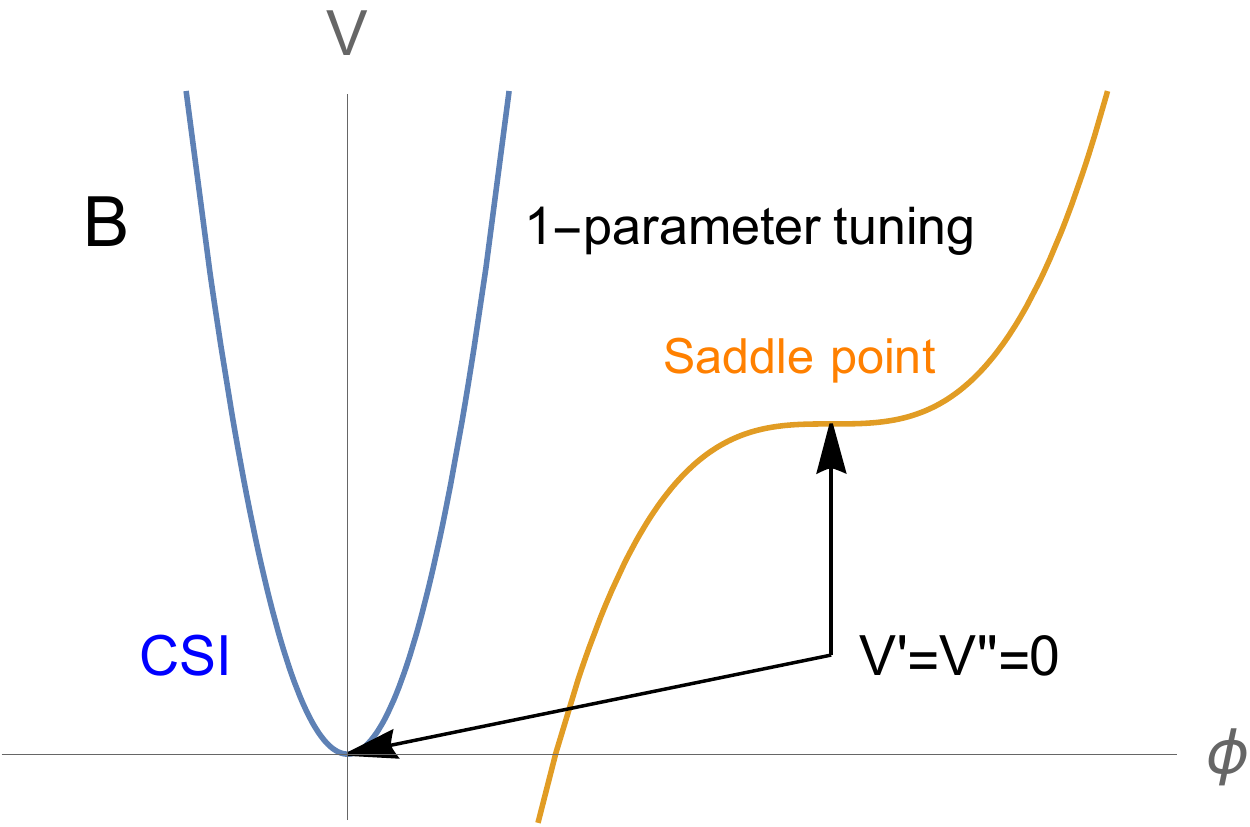}
\includegraphics[width=7cm]{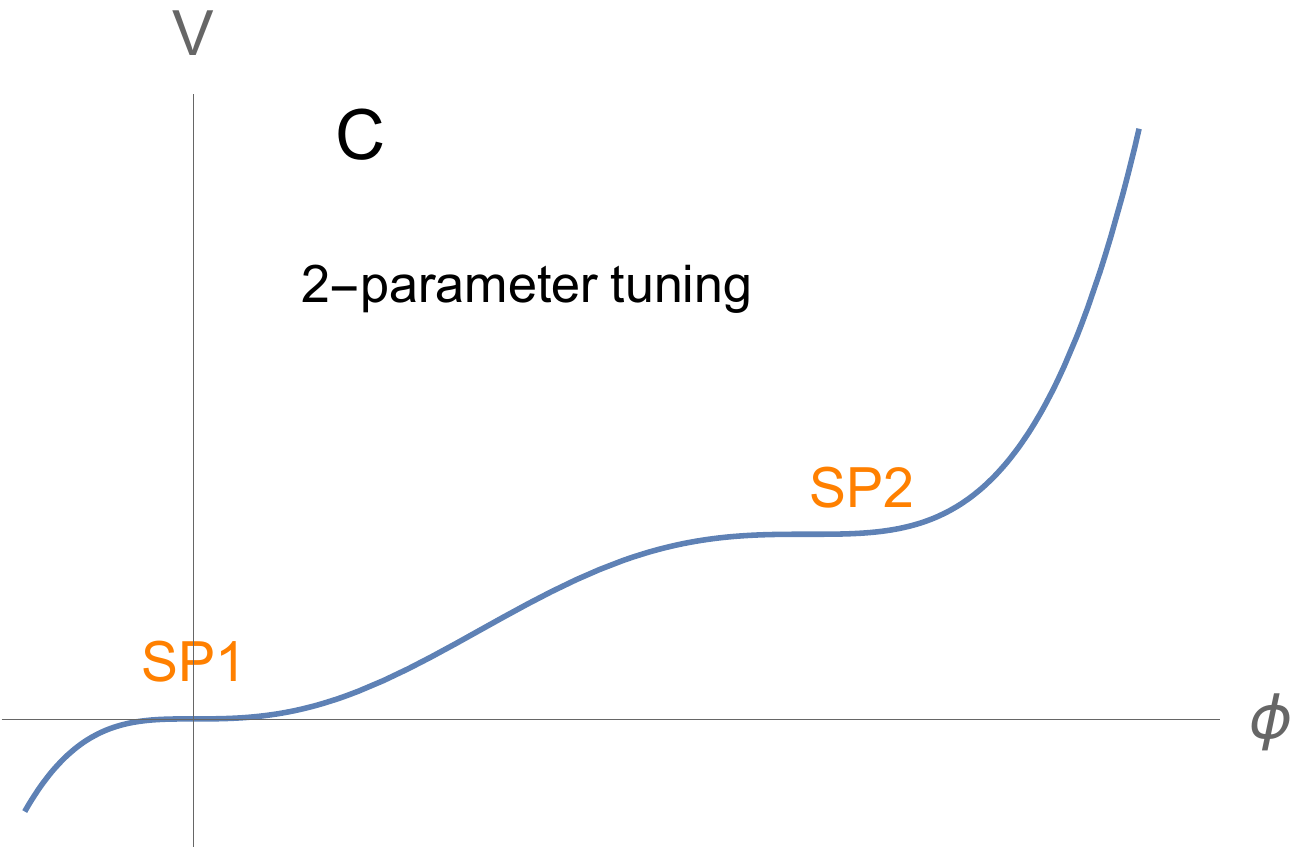}
\caption{Typical examples of the effective potential resulting from the MPP.}
\label{schematic}
\end{center}
\end{figure}

Here we briefly review the consequence of the MPP, focusing on what kind of effective potentials  the MPP can result in. Before proceeding, we first note that the effective potential exists independently of the renormalization scale, up to the renormalization of the field, for a given set of bare parameters. (In other words, the renormalization scale dependence should only arise due to truncation at some loop level.) In Fig.~\ref{schematic}, we list examples of the possible effective potentials resulting from the MPP:
In panel A, we show the first kind of tuning $V_{\phi=\phi_{\rm ours}}^{}=V_{\phi=\phi_{\rm another}}^{}$ given in the original version of the MPP~\cite{Froggatt:1995rt,Froggatt:2001pa,Nielsen:2012pu}. 
 In panel B, we show another one-parameter tuning $V'=V''=0$, which provides an explanation for the CSI-type potential $\left.V'\right|_{\phi=0}=\left.V''\right|_{\phi=0}=0$. This is also the key assumption for the Coleman-Weinberg mechanism.
From this point of view, the CSI-type potential is just one of the possible critical points, in the theory space, in the MPP.
In panel C, we show the potential (called CP 2-2) that is a consequence of two-parameter tuning to realize two saddle points $V'|_\tx{SP1}=V''|_\tx{SP1}=0$ and $V'|_\tx{SP2}=V''|_\tx{SP2}=0$. We stress that all the above potentials are achieved in the context of the MPP irrespectively of any kind of scale invariance. Therefore, it does not matter whether other fields such as the right-handed neutrinos have mass parameters in the action or not.

Among the various critical points analyzed so far, we take the CP 2-2 defined in Ref.~\cite{Hamada:2020wjh} that has two saddle points in the effective potential, because the widest parameter region is allowed by the constraints considered. 
The renormalized Lagrangian is
\aln{
{\cal L}={\cal L}_{\rm SM} &-\frac{1}{2}(\partial_\mu \phi)^2-\frac{1}{2}(\partial_\mu S)^2
-\lambda_H (H^\dagger H)^2
-\frac{\lambda_\phi }{4!}\phi^4-\frac{\lambda_{\phi S} }{4}\phi^2S^2-\frac{\lambda_S }{4!}S^4+\frac{\lambda_{\phi H} }{2}\phi^2(H^\dagger H)
\nn
&-\frac{\lambda_{SH} }{2}S^2(H^\dagger H)
-\frac{\mu_\phi }{3!}\phi^3
+\frac{1}{2}\sum_{i=1}^3\overline{\nu_{Ri}}\gamma^{\mu}i\partial_\mu \nu_{Ri} -\frac{1}{2}\sum_{i=1}^3 M_{Ri} \overline{{\nu_{Ri}}^{c}}\nu_{Ri} 
\nn
&
 -\sum_{i,j=1}^3\left( y_{\nu ij} \overline{L_i}H^c \nu_{Rj}  + \frac{y_{ij}^{\phi}}{2}\phi\, \overline{{\nu_{Ri}}^{c}}\nu_{Rj} +{\rm h.c} \right)~, 
\label{lagrangian}
}
where ${\cal L}_{\rm SM} $ is the SM Lagrangian without the Higgs potential and without the right-handed neutrinos, and $H^c \equiv i\sigma_2H^*$. 
Note that all the couplings in the Lagrangian are defined by the  zero-momentum subtraction scheme.   
These couplings are the same as those in the effective potential. 
It is precisely these couplings that are restricted by the MPP. 
Here, among several realizations of the MPP, we choose a   class of simple solutions that satisfy the vanishing 
of all the terms with dimensionful couplings except for $\phi^3$, namely, $\phi$, $\phi^2$, $S^2$, $H^\dagger H$, $\phi H^\dagger H$, and $\phi S^2$. 
In Eq.~(\ref{lagrangian}), we have omitted these terms from the beginning. 
Note that the renormalized mass of $S$ around the origin $h=\phi=S=0$ is set to zero by the MPP, while it obtains the finite mass around the true vacuum as we will see in Eq.~(\ref{eq:dm mass}).  


%

For simplicity, we will make the following three assumptions. 
First, we take $\lambda_S $ to be zero at the EW scale since it is irrelevant for our discussion.\footnote{
This choice makes the perturbativity bound loosest, while keeping the stability of the effective potential.
}
Second, we assume that the Majorana Yukawa couplings $y_{ij}^{\phi}$ are not large and do not affect the other RG runnings.\footnote{
If one wants, one can forbid it by a $Z_2'$ symmetry $\phi\rightarrow-\phi$ that is softly broken by the $\phi^3$ term.   
} 
Third, we assume that Majorana masses are all degenerate each other, $M_{R1} =M_{R2} =M_{R3} =M_R $, and neglect the mixing in the neutrino Dirac Yukawa couplings, i.e., $y_{\nu ij} =y_{\nu i}\delta_{ij}$. 
%
At the leading order of $M_R^{-1}$, the neutrino mass matrix becomes
\aln{
(m_\nu )_{ij} 
=	{v^2\over2M_R}\delta_{ij}y_{\nu i}^2.
} 
We call the number of ``heavier active neutrinos'' $n_\nu $, namely the degeneracy of the largest~$y_{\nu i} $:
In the case of normal and inverted hierarchies, we have $n_\nu =1$ and $2$, respectively, and
when all the neutrino masses are degenerate, we have $n_\nu =3$.
Throughout this paper, we will use $m_\nu =0.05~{\rm eV}$~\cite{PDG2020} as the heaviest mass of active neutrinos.\footnote{
The current upper limit of the sum of the neutrino masses is given by the Planck and baryon acoustic oscillation measurements as
$
\sum m_\nu <0.12\ {\rm eV}\quad (95 \%~\text{CL})
$~\cite{Aghanim:2018eyx}, 
which corresponds to $m_{\nu} \sim 0.04\ {\rm eV}$ in the degenerate case. Although it is ruled out, we also show the $n_\nu=3$ case in this paper to illustrate the $n_\nu$ dependence.
}
The relation between $M_R $ and the largest $y_\nu $ is then determined to be
$y_\nu \simeq 0.41\sqrt{M_R /10^{14}\ {\rm GeV}}$.
%

Here, we briefly review how the CW mechanism realizes the EW symmetry breaking in our model with the MPP. 
The effective potential for the $\phi$--$H$ system can be expressed at one-loop level as
\begin{align}
V_{\rm eff} & = \lambda_H (H^\dagger H)^2 + \frac{\lambda_{\phi H}}{2} \phi^2(H^\dagger H)  +\frac{\mu_\phi }{3!}\phi^3 +\frac{\lambda_\phi }{4!}\phi^4  \notag\\
&+ \frac{M_\phi^4(\phi)}{64\pi^2}\left[ \ln\frac{M_\phi^2(\phi)}{\mu^2} -\frac{1}{2}\right]  + \frac{\lambda_{\phi S}^2 \phi^4}{256\pi^2}\left[ \ln\frac{\lambda_{\phi S} \phi^2}{2\mu^2} -\frac{1}{2}\right] + \Delta V_{\rm 1-loop}(h,\mu), 
\end{align}
where $M_\phi^2(\phi) = \mu_\phi \phi + \lambda_\phi \phi^2$, and the renormalization scheme applied in \cite{Haruna:2019zeu} is used.  
The $\Delta V_{\rm 1-loop}$ term is the 1-loop potential for $h$ given in Eq.~(\ref{one-loop Higgs potential}). 
In this expression, we assume that the $H$--loop contribution to the $\phi^4$ term as well as the 
field dependent masses of $\phi$ and $S$ coming from $H$ are negligibly small, which can be realized by taking 
$\lambda_{\phi H} \ll \lambda_{\phi S}$ and $\lambda_{SH}H^\dagger H \ll \lambda_{\phi S}\phi^2$. 
This potential can be rewritten at the scale $\mu = \mu_*$ where $\lambda_\phi$ vanishes as 
\begin{align}
V_{\rm eff} & = \lambda_H \left(H^\dagger H - \frac{\lambda_{\phi H}}{\lambda_H}\phi^2\right)^2  
+ \Delta V_{\rm 1-loop}(h,\mu_*)  \notag\\
& +\frac{\mu_\phi }{3!}\phi^3+ \frac{\mu_\phi^2\phi^2}{64\pi^2}\left[ \ln\frac{\mu_\phi\phi}{\mu_*^2} -\frac{1}{2}\right] + 
\frac{\lambda_{\phi S}^2 \phi^4}{256\pi^2}\left[ \ln\frac{\lambda_{\phi S} \phi^2}{2\mu_*^2} -\frac{1}{2}\right] -\frac{\lambda_{\phi H}^2}{\lambda_H}\phi^4, \label{eq:phi_h}
\end{align}
In the following, we also assume that the $\mu_\phi^2\phi^2$ term is much smaller than the $\lambda_{\phi S}^2 \phi^4$ term.~\footnote{Note also that other scalar couplings such as $\phi |H|^2$ and $\phi S^2$, which are set to zero by the MPP, are induced at the one-loop level as well. 
However, in the present analysis, the mixing between $H$ and $\phi$ is small, and since  we are interested in $\langle S\rangle =0$, these effects are small.  
}
Requiring the existence of the two saddle points at $\phi = 0$ and $\phi = \phi_{\rm saddle}$ (CP2-2)\footnote{Note that this saddle point is nothing to do with the saddle point of the Higgs potential discussed in the next section.}, $\mu_\phi$ is determined to be 
\begin{align}
\mu_\phi=\frac{\lambda_{\phi S}^2}{16\pi^2}\phi_{\rm saddle}, \quad \text{with}\quad \phi_{\rm saddle} = -\frac{1}{e}\sqrt{\frac{2}{\lambda_{\phi S}}}\mu_*e^{\frac{8\pi^2 \lambda_{\phi H}^2}{\lambda_H\lambda_{\phi S}^2}}. 
\label{trilinear}
\end{align}
The vacuum expectation value (VEV) of $\phi$, $\langle\phi\rangle$, is then determined as 
\begin{align}
\langle\phi\rangle = e^{W(1/e)}(-e\phi_{\rm saddle}), \label{eq:phivev}
\end{align}
where $W$ is the Lambert $W$ function. 
Using Eqs.~(\ref{trilinear}) and (\ref{eq:phivev}), we obtain the relation between $\mu_*$ and $\langle\phi\rangle $ as 
\begin{align}
\mu_* = \sqrt{\frac{\lambda_{\phi S}}{2}}\langle\phi\rangle \exp \left[-W(e^{-1}) - \frac{8\pi^2\lambda_{\phi H}^2}{\lambda_H^{}\lambda_{\phi S}^2}\right].  
\label{mu star}
\end{align}
Now, it is clear that by looking at the first line of Eq.~(\ref{eq:phi_h}) the EW symmetry breaking is triggered by $\langle\phi \rangle$:  
\aln{
\frac{v}{\langle\phi\rangle}=\sqrt{\frac{\lambda_{\phi H} }{2\lambda_H }}~,
\label{ew scale}
}
where $v\simeq 246$ GeV.
At the minimum, the squared masses of $\phi$ and $S$ are given by
\aln{
m_\phi^2\sim -\frac{\lambda_{\phi H}^{}}{2}v^2+\frac{1+W(e^{-1})}{32\pi^2}\lambda_{\phi S}^{2}\langle \phi\rangle ^2~,
\quad m_S^2 = \frac{\lambda_{SH}}{2} v^2+\frac{\lambda_{\phi S}}{2}\langle\phi\rangle^2~,
\label{eq:dm mass}
}
where we have neglected the mixing between $h$ and $\phi$. 
From the above discussion, the six parameters $\lambda$, $\lambda_\phi$, $\lambda_{\phi S}$, $\lambda_S$, $\lambda_{\phi H}$, and $\lambda_{SH}$ in the potential are reduced by fixing the two parameters, $v$ and the Higgs mass $125.1$\,GeV, and by taking $\lambda_S$ to be zero at the EW scale as aforementioned. The resultant three parameters are
\aln{
m_S ,\quad
\lambda_{SH},\quad
\langle \phi \rangle,
}
where $\lambda_\phi$ has been replaced by the scale $\mu_*$ at which $\lambda_\phi$ vanishes, and further converted to~$\langle \phi \rangle$ by Eq.~\eqref{mu star}.

\begin{figure}[t!]
\begin{center}
\includegraphics[width=10cm]{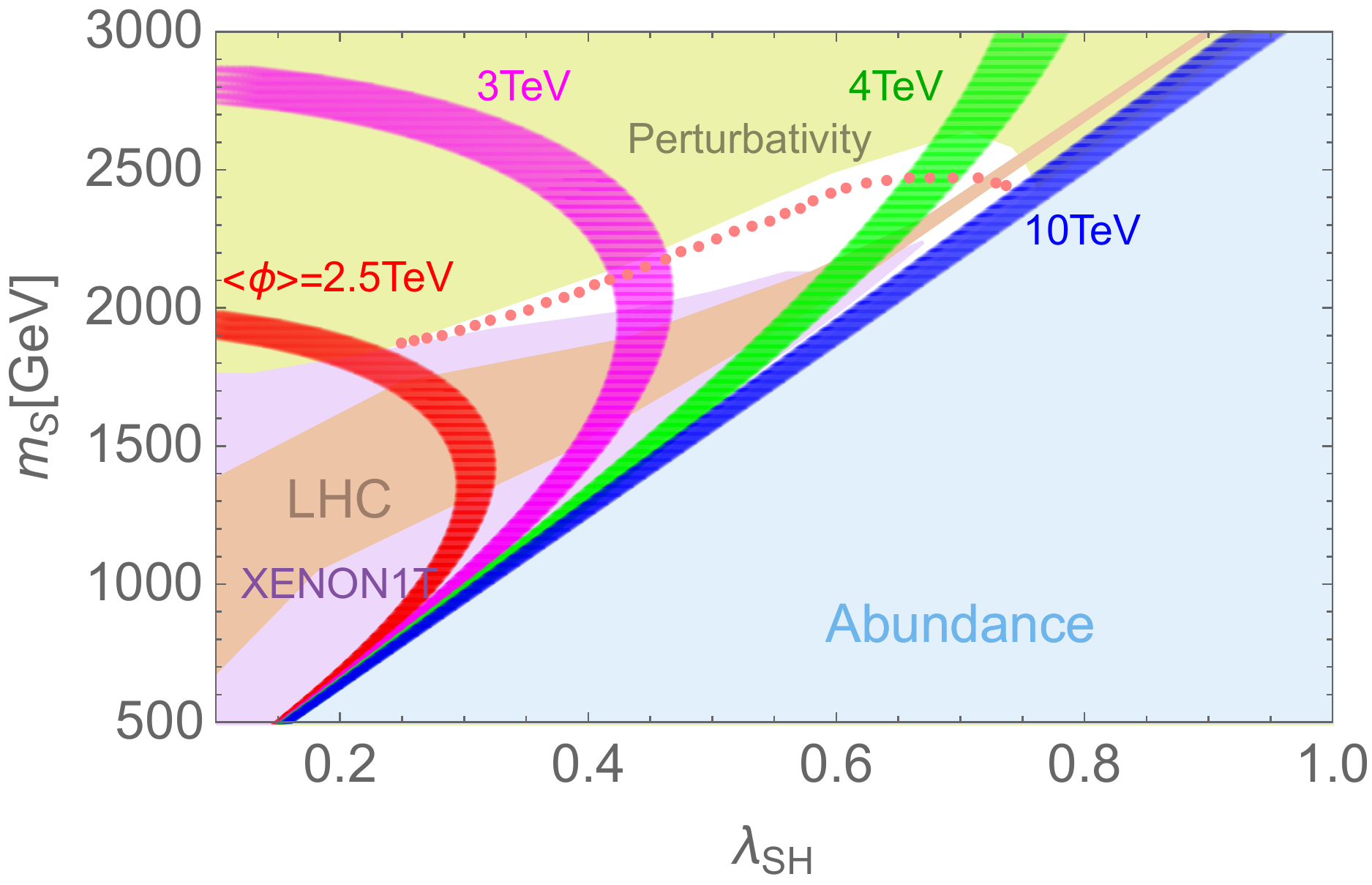}
\caption{
The solid curves correspond to Eq.~(\ref{fitting}) for $\langle\phi\rangle =2.5$ TeV (red), 3 TeV (magenta), 4 TeV (green), and 10 TeV (blue), with each width corresponding to the error of $m_\text{th}=1590\pm40$ GeV.
Shaded regions are respectively excluded by the XENON1T experiment (purple), the LHC data (orange), DM abundance (cyan), and the perturbativity bound (yellow).
Regarding the perturbativity bound, the absence of Landau pole up to $\mu=10^{17}$ GeV is imposed.  
}
\label{fig:CP22}
\end{center}
\end{figure}

The thermal relic abundance of $S$ is satisfied when~\cite{Hamada:2020wjh}
\aln{
&4\lambda_{SH}^2+\lambda_{\phi S}^2=\left(\frac{m_S }{m_{\rm th} }\right)^2,\quad m_{\rm th} =1590\pm40~\text{GeV}. 
\label{fitting}
} 
This relation provides a contour of $\langle\phi\rangle$ in the $m_S$ vs $\lambda_{SH}$ plane as shown in Fig.~\ref{fig:CP22}:
The red, magenta, green, and blue curves correspond to $\langle\phi\rangle=2.5$ TeV, $3$ TeV, $4$ TeV, and $10$ TeV, respectively.
We also show the other constraints on the model parameters in the figure:
The purple, orange, cyan, and yellow shaded regions are excluded by the updated XENON1T result~\cite{Aprile:2018dbl},\footnote{
For the region with the DM mass larger than 1 TeV, we extrapolate the 
upper limit on the spin independent cross section for DM and nucleon scatterings.} the LHC results, DM relic abundance, and perturabativity bound,
respectively~\cite{Hamada:2020wjh}.
We note that these bounds are insensitive to $M_R$.\footnote{
$M_R$ might possibly affect the perturbativity bound via the RGE of $\lambda_S$ and $\lambda_{SH}$, but its effect is only through the small $\lambda_H$ coupling and is negligible.
}

In the following analyses of the critical Higgs inflation, we choose the parameters on the dotted line in Fig.~\ref{fig:CP22} that is close to the perturbativity bound, to reduce the computational time.\footnote{
When we vary $m_S$ from the dotted line for a fixed $\lambda_{SH}$, the coupling $\lambda_{\phi S}$ changes via Eq.~\eqref{eq:dm mass}, but the running of $\lambda$ and $\lambda_H$ do not depend on $\lambda_{\phi S}$ at the one-loop level (see Eq.~\eqref{lambda H beta function}), and hence the high-scale Higgs potential is not altered drastically. 
Note also that, even though the dotted line is close to the perturbativity bound, $\lambda_{SH}$ (which contributes to running of $\lambda$ and $\lambda_H$ at one-loop) remains perturbative unlike $\lambda_S$, and hence the effective Higgs potential is reliably computed.
}
On the dotted line, there remains only single parameter in the scalar sector, and we choose it to be $\lambda_{SH}$.
Therefore, there are two parameters $\lambda_{SH}$ and $M_R$ in total.
We choose these parameters in such a way that the Higgs potential has a (near) saddle-point at high scale region.\footnote{
Without the right-handed neutrinos, the high-scale potential value tends to become too large to accommodate the observed value of the tensor-to-scalar ratio; see also Ref.~\cite{Hamada:2017sga}.
}

\section{Saddle point of Higgs potential}\label{saddle point section}
Now that the TeV-scale couplings are obtained, we extrapolate them towards high scales.
In this section, we analyze the effective Higgs potential for large field values, and look for its saddle point in order to realize the critical Higgs inflation.

\subsection{Effective potential}\label{effective potential section}
We calculate the one-loop effective Higgs potential improved by the two-loop RGEs presented in \ref{RGEs}.
The one-loop effective potential for large $h$ in the $\overline{\rm MS}$ scheme in the Landau gauge is 
\aln{
V_{} &=\frac{\lambda_H (\mu)}{4}\bar{h}^4+\Delta V_\text{1-loop}(h,\mu),
	\label{full 1-loop potential}
}
where
\aln{
\Delta V_\text{1-loop}(h,\mu)
	&=	\frac{6M_{W}^4(h)}{64\pi^2}\left[\ln \left(\frac{M_{W}^2(h)}{\mu^2}\right)-\frac{5}{6}\right]+\frac{3M_{Z}^4(h)}{64\pi^2}\left[\ln \left(\frac{M_{Z}^2(h)}{\mu^2}\right)-\frac{5}{6}\right]\nn
	&\quad
		-\frac{3M_{t}^4(h)}{16\pi^2}\left[\ln \left(\frac{M_{t}^2(h)}{\mu^2}\right)-\frac{3}{2}\right]
	-\sum_{i=1}^3 \frac{M_{N_i}^4(h)}{32\pi^2}\left[\ln \left(\frac{M_{N_i}^2(h)}{\mu^2}\right)-\frac{3}{2}\right]
	\nn
	&\quad
		+\frac{M_S^{4}(h)}{64\pi^2}\left[\ln \left(\frac{M_S^2(h)}{\mu^2}\right)-\frac{3}{2}\right]+\frac{M_\phi^{4}(h)}{64\pi^2}\left[\ln \left(\frac{M_\phi^2(h)}{\mu^2}\right)-\frac{3}{2}\right],
\label{one-loop Higgs potential}
}
in which the effective masses are
\aln{
M_W (h)
	&=	\frac{g_2 h}{2},&
M_Z (h)
	&=	\frac{\sqrt{g_2^2+g_Y^2}}{2}h,&
M_t (h)
	&=	\frac{y_t h}{\sqrt{2}},\\
M_{N_i} (h)
	&=	\frac{M_R }{2}\left(1+\sqrt{1+\frac{2y_{\nu i}^2 h^2}{M_R^2}}\right),&
M_S^2(h)
	&=	
\frac{\lambda_{SH} }{2}h^2,
&
M_\phi^{2}(h)
	&=	\frac{\lambda_{\phi H} }{2}h^2,
}
and
\aln{
\overline{h}
	:=he^{\Gamma(\mu)}=	h\exp\left(\int_{0}^{\ln \mu/M_t }ds\,\gamma_H \right)
}
is the Higgs field with field renormalization. 
See Eq.~(\ref{gammaH}) in \ref{RGEs} for the one-loop result of $\gamma_H^{}$.   
We here represent $N_i$ as the mass eigenstates for the heavy Majonara neutrinos.  
We neglect the contributions from the leptons and quarks other than the top quark because their Yukawa couplings are small.
We also neglect the loops of the Higgs and the NG bosons because $\lambda_H $ becomes small at high scales.

In the following, we match two theories with and without right-handed neutrinos around $\mu=M_R $, to obtain the threshold correction.
We expand the one-loop correction from heavy neutrinos by $h$ as 
\aln{
&-\sum_{i=1}^3 \frac{M_{N_i}^4(h)}{32\pi^2}\left[\ln \left(\frac{M_{N_i}^2(h)}{\mu^2}\right)-\frac{3}{2}\right]
\nn
&=	-\frac{1}{32\pi^2}\Bigg[\frac{3M_R^4}{2}\left(-3+4\ln\left(\frac{M_R}{\mu}\right)\right)+2h^2M_R^2n_\nu y_{\nu}^2\left(-1+2\ln \left(\frac{M_R }{\mu}\right)\right)
\nn
&\phantom{=	-\frac{1}{32\pi^2}\Bigg[}
	+\frac{h^4}{2}n_\nu y_{\nu}^4\left(1+2\ln\left(\frac{M_R }{\mu}\right)\right)+\cdots 
\Bigg],
	\label{expanded threshold correction}
}
where the coefficient of $h^4$ corresponds to the threshold correction to $\lambda_H$ below $M_R $ \cite{Burgess:2007pt,Iso:2018aoa,Bando:1992wy}:\footnote{
The coefficient of $h^2$ in Eq.~\eqref{expanded threshold correction} gives the threshold correction to the Higgs mass-squared parameter in the low-energy theory. 
At the CP 2-2, the mass-squared parameter including this correction is tuned to zero, based on the MPP. 
See Refs.~\cite{
Haruna:2019zeu,Hamada:2020wjh} 
for the detailed discussion. 
}
\aln{
\Delta\lambda_{H}^{(R)}
	:=	
		-\frac{n_\nu y_\nu^4}{16\pi^2}\left(1+2\ln \left(\frac{M_R }{\mu}\right)\right). 
}
The term containing $\ln(M_R /\mu)$ leads to the subtractions of $y_\nu^4$ term from the beta function.  
The same analysis can be also applied to the field renormalization as 
\aln{
\frac{1}{2}h\Box h\left[1-\frac{2}{32\pi^2}\sum_{i=1}^3 y_{\nu i} \ln\left(\frac{{M_{N_i}}^2(h)}{\mu^2}\right)\right]\simeq \frac{1}{2}h\Box h\left[1-\frac{1}{8\pi^2}n_\nu y_\nu^{2}\ln\left(\frac{M_R }{\mu}\right)+\cdots\right],
}
from which we can see that the canonically normalized Higgs field $h_c $ in the low-energy theory is given by
\aln{
h_c 
	=	h\left[1-\frac{1}{8\pi^2}n_\nu y_\nu^{2}\ln\left(\frac{M_R }{\mu}\right)\right]^{1/2}
	=:	hZ_R . 
}
This also leads to the redefinition of the quartic coupling. 
By combing both of them, the Higgs quartic coupling below $\mu=M_R $ is\footnote{
If we want, we may take into account the threshold correction of $S$ too:
$\Delta\lambda_H^{(S)}=\frac{\lambda_{SH}^2}{16\pi^2}\left(\frac{1}{6}+\frac{1}{2}\ln\left(\frac{m_S }{\mu}\right)\right)$. This contribution is minor for our analysis and we neglect it hereafter. 
}
\aln{
\lambda
	&:=	\lambda_H Z_R^{-4}+\Delta\lambda_H^{(R)}
\nn
	&=	\lambda_H \left(1-\frac{4n_\nu }{16\pi^2}y_\nu^{2}\ln\left(\frac{\mu}{M_R }\right)
\right)+\frac{n_\nu y_\nu^4}{16\pi^2}\left(-1+2\ln \left(\frac{\mu}{M_R }\right)\right). 
\label{relation of quartic coupling}
}
One can easily check that contributions from heavy neutrinos cancel out in the beta function ${d\lambda/d\ln\mu}$.

\begin{figure}[t!]
\begin{center}
\includegraphics[width=0.48\textwidth]{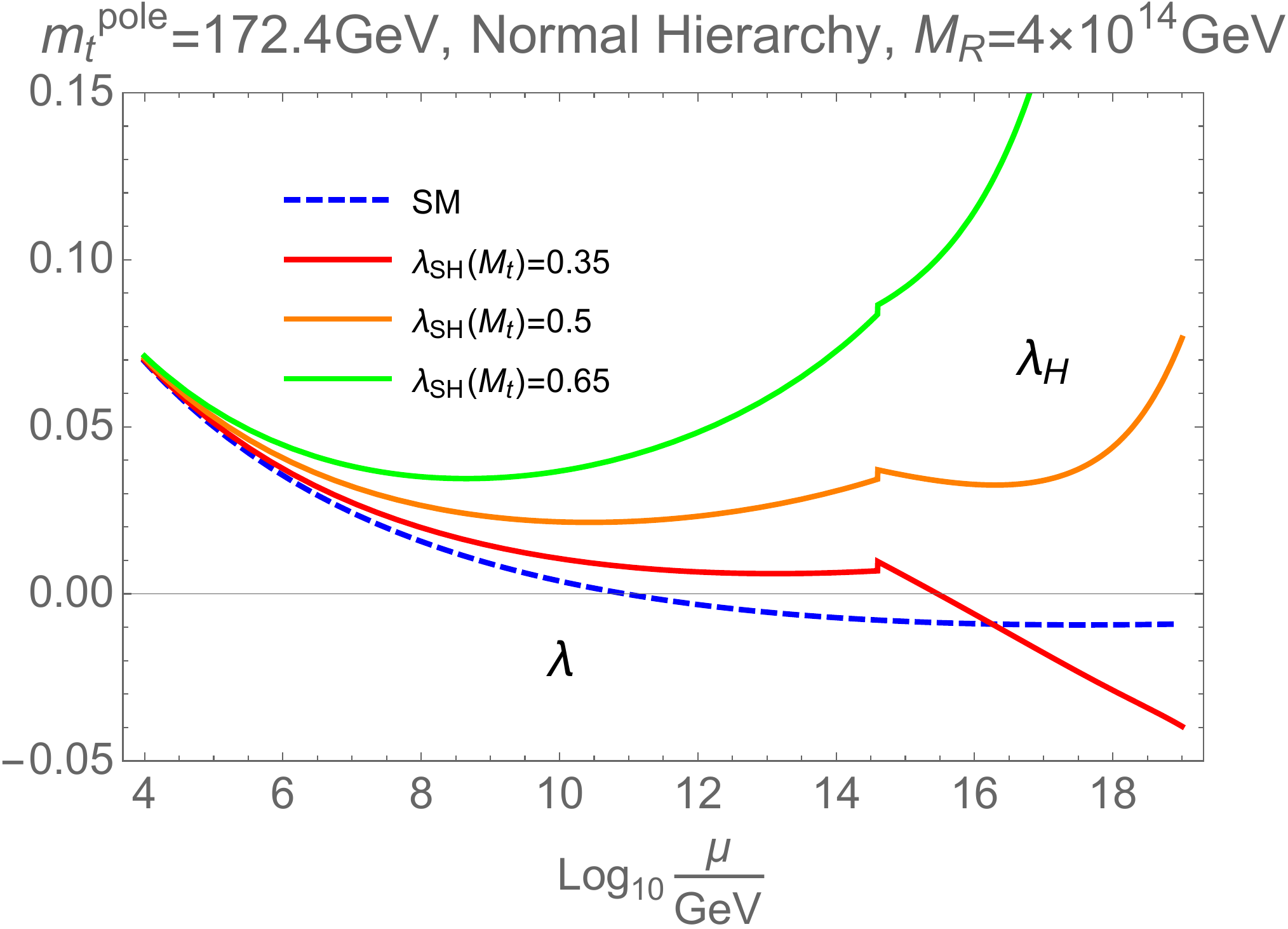}
\includegraphics[width=0.48\textwidth]{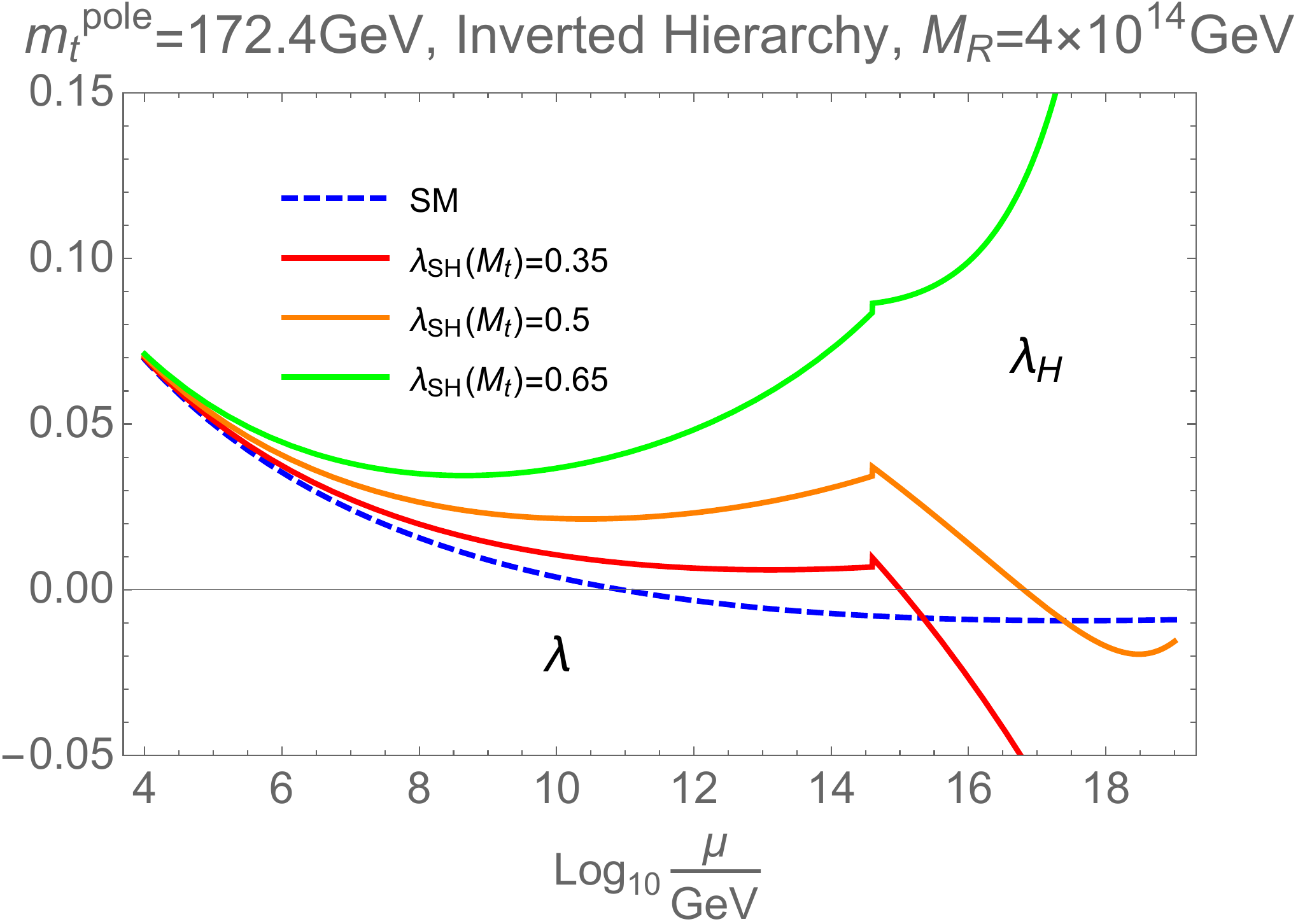}
\includegraphics[width=0.48\textwidth]{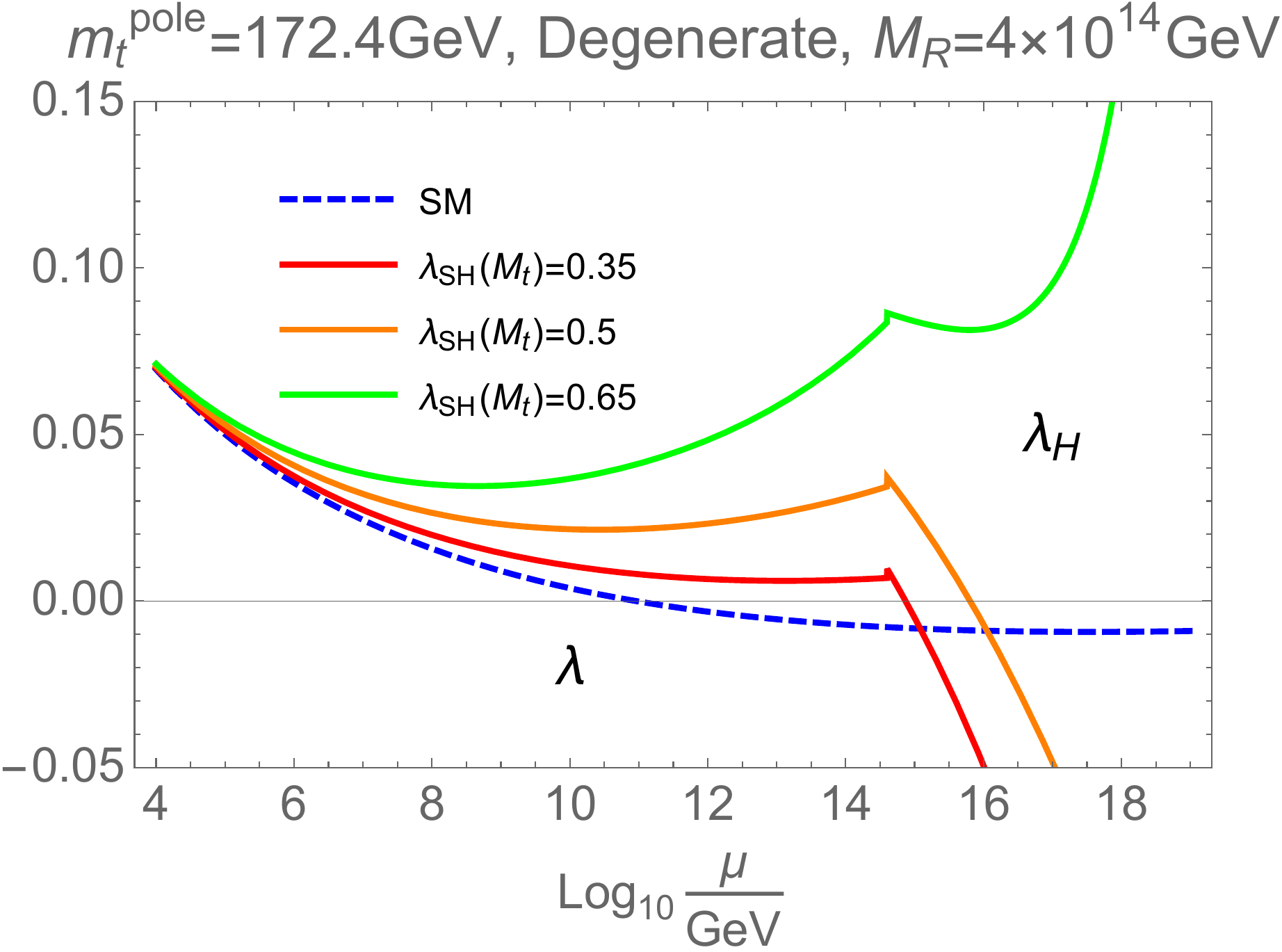}
\caption{
The RG runnings of $\lambda$ and $\lambda_{H} $. 
Upper left (right) corresponds to the case of normal (inverted) hierarchy and lower panel corresponds to the degenerate case.  
Here, the different colors correspond to the different values of $\lambda_{SH} (M_t )$. 
}
\label{fig:lambda}
\end{center}
\end{figure}

In Fig.~\ref{fig:lambda}, we show the RG runnings of $\lambda$ and $\lambda_H $, where the upper left (right) panel corresponds to the normal (inverted) hierarchy case and the lower panel corresponds to the degenerate case.  
The different colors correspond to the different values of $\lambda_{SH} (M_t )$, and the top pole mass $m_t^\text{pole}$ and $M_R $ are fixed at $m_t^\text{pole}=172.4$ GeV \cite{PDG2020} and $4\times 10^{14}$ GeV, respectively. 
As explained above, other parameters, $\langle\phi\rangle$ and $m_S $, are fixed as functions of $\lambda_{SH} $ on the dotted line in Fig.~\ref{fig:CP22} by the thermal relic abundance of $S$ to explain $\Omega_{\rm DM} h^2\sim 0.12$. 
%

\subsection{Saddle point}\label{sec:saddle}
In the following, we rewrite the one-loop Higgs effective potential~(\ref{full 1-loop potential}) as 
\aln{
V_{} =\frac{\lambda_{\rm eff} (h,\mu)}{4}h^4, \label{effective Higgs potential}
}
where
\aln{\lambda_{\rm eff} (h,\mu)=\left[
	\lambda_H (\mu)e^{4\Gamma(\mu)}
	+4{\Delta V_\text{1-loop}(h,\mu)\over h^4}
	\right].  
}
Note that $\lambda_\text{eff}(h,\mu)$ is independent of the renormalization point $\mu$ if we take all order loop corrections into account. For practical purpose with finite-loop, it better approximates by choosing $\mu\sim h$.

In the SM without the right-handed neutrinos, it is known that $\lambda_\text{eff}(h,h)$ in general takes a minimum value $\lambda_\text{min}$ at $h=h_\text{min}$ ($\simeq 10^{18}$ GeV); see the blue dashed line in Fig.~\ref{fig:lambda}.
Furthermore, by tuning the top mass, we can realize a saddle point $V_{}'=V_{}''=0$ at $h=h_s$ ($\simeq h_\text{min}$)~\cite{Hamada:2014wna}. (This tuning may be done within the 1.4$\sigma$ experimental bound as stressed in Introduction.) However, this saddle-point itself is not sufficient to achieve a viable saddle-point inflation because the resultant CMB fluctuations become too large~\cite{Isidori:2007vm,Hamada:2013mya,Fairbairn:2014nxa}.
By introducing the non-minimal coupling $\xi |H|^2 R$, a successful inflation can be achived around the (near) saddle-point even when $\xi\sim 10$~\cite{Hamada:2014iga,Bezrukov:2014bra,Hamada:2014wna}. 
This is called the critical Higgs inflation.

In this paper, we pursue the critical Higgs inflation in our model at the CP 2-2.    
The detailed analysis will be presented in the next section. 
In the remaining of this section, we look for the saddle point in our model.
When we take into account the right-handed neutrinos, they first lower $\lambda_H$ above $M_R$, and at higher scales, the extra scalar couplings become large and their contributions raise $\lambda_H$ again. As a result, we may have a minimum for $\lambda_H$ at a high scale as can be seen in Fig~\ref{fig:lambda}, e.g.\ with the orange curve for the normal and inverted hierarchy cases, and green for the degenerate case.

In general, the condition to have a saddle point is
\aln{
V_{}^{'}=V^{''}
	&=	0 &
&\Leftrightarrow&
\lambda_{\rm eff}(h,h) +\frac{1}{4}{d\lambda_{\rm eff}(h,h)\over d\ln h}=12\lambda_{\rm eff}(h,h) + 8{d\lambda_{\rm eff}(h,h)\over d\ln h}+{d^2\lambda_{\rm eff}(h,h)\over {d(\ln h)^2}}
	&=	0.
\label{saddle condition} 
}
Again, we call the position of saddle-point $h_s^{}$. 
As said in the last of Sec.~\ref{model}, there is only single parameter $\lambda_{SH}$ in the scalar sector.
For each given $\lambda_{SH}$, we numerically solve for $M_R$ that achieves the saddle-point condition~\eqref{saddle condition}, and obtain its position $h_s$.

\begin{figure}[t!]
\begin{center}
\includegraphics[width=0.48\textwidth]{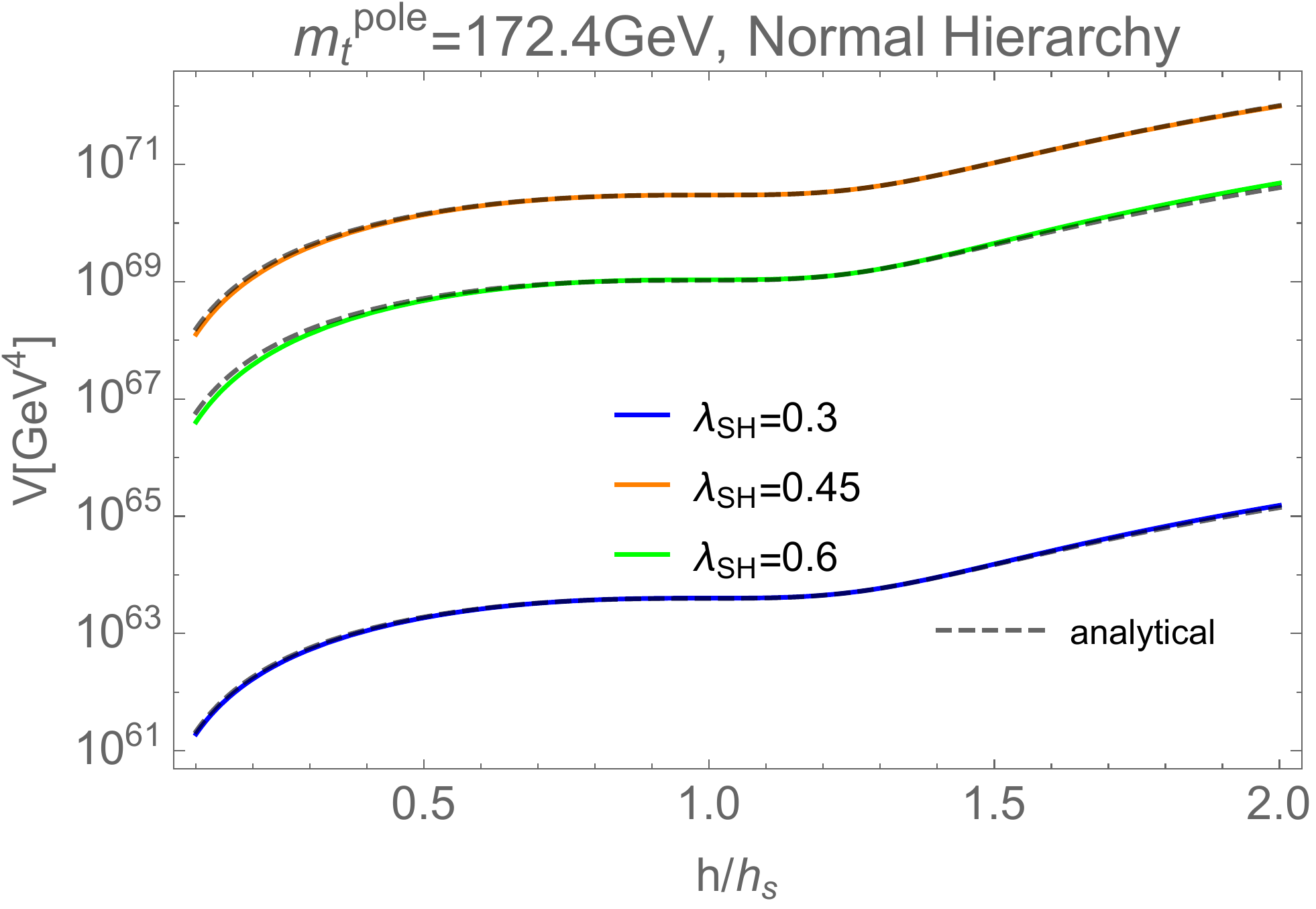}
\includegraphics[width=0.48\textwidth]{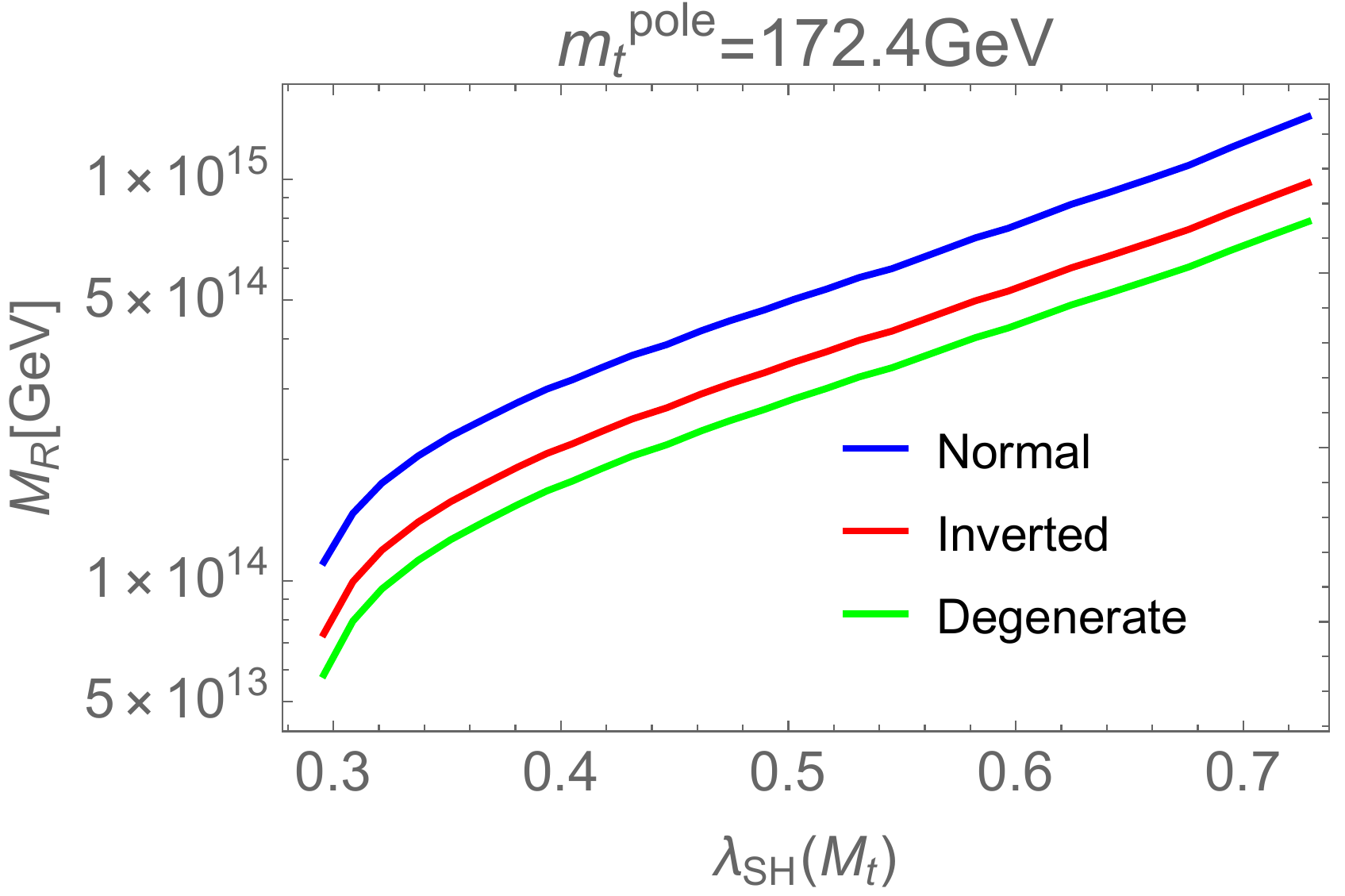}
\\
\includegraphics[width=0.48\textwidth]{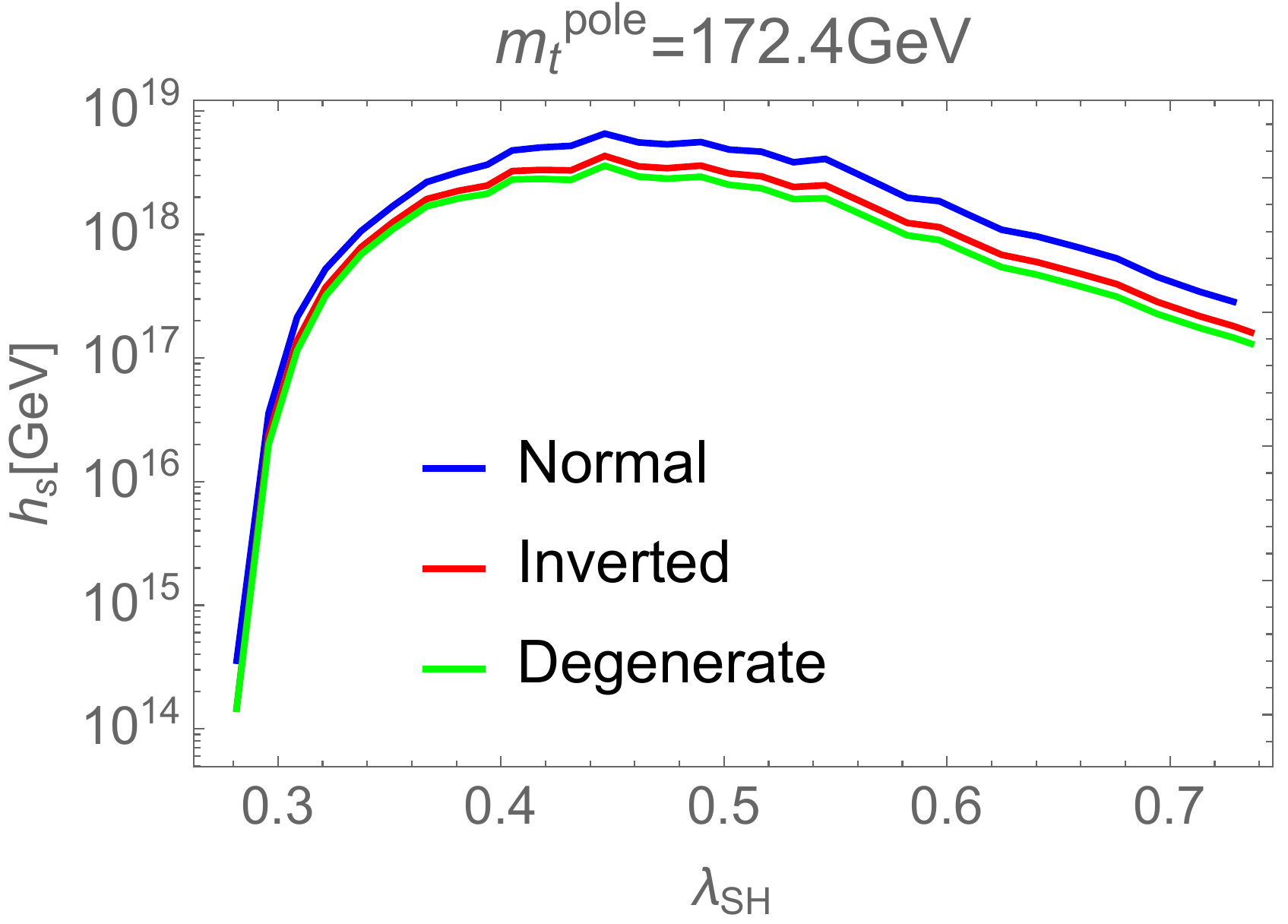}
\includegraphics[width=0.48\textwidth]{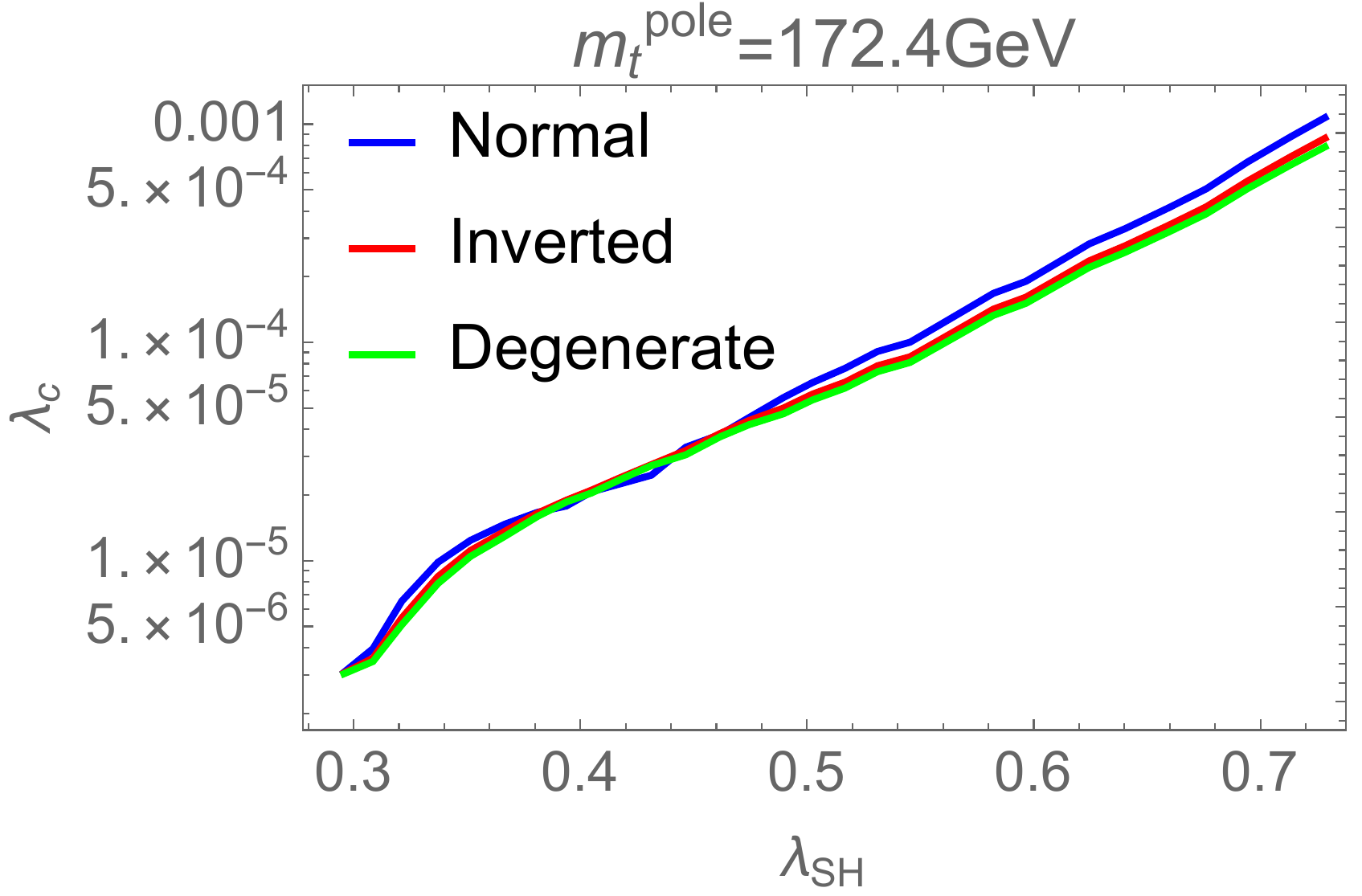}
\caption{Upper Left: The one-loop effective Higgs potential in the normal hierarchy case where the different colors correspond to the different values of $\lambda_{SH} (M_t )$. 
Here, dashed black curves represent the Higgs potential whose quartic coupling is replaced by the analytical one~(\ref{approximate lambda}). 
\\
Upper Right: The relation between $M_R $ and $\lambda_{SH} (M_t )$ determined by the saddle point condition~(\ref{saddle condition}). 
\\
Lower Left: The saddle point $h_s $ as a function of $\lambda_{SH} (M_t )$, where blue and red correspond to the normal and inverted hierarchy, while green to degenerate.     
\\
Lower Left: The minimum value of $\lambda_H $ as a function of $\lambda_{SH} (M_t )$.   
}
\label{fig:saddle}
\end{center}
\end{figure}
In the upper left panel in Fig.~\ref{fig:saddle}, we show the Higgs potential in the case of the normal hierarchy where the differently colored contours  correspond to the different values of $\lambda_{SH} $. 
Here, the Higgs field $h$ is normalized by $h_s $. 
In the upper right panel, we plot the value of $M_R $ as a function of $\lambda_{SH} $ that is determined by the saddle point condition~(\ref{saddle condition}). 
In the lower left panel, we show $h_s$ as a function of $\lambda_{SH} $ where blue and red correspond to the normal and inverted hierarchy cases, respectively, while green to degenerate. 
One can see that $h_s $ has a strong dependence on $\lambda_{SH} $, which causes the strong dependence of the value of the potential around the saddle point on $\lambda_{SH} $ in the upper left panel. 

The above numerical result can be fitted by an analytic formula as follows.
Around the minimum $h=h_\text{min}$ giving the minimal value of $\lambda_\text{eff}$, we may always approximate as~\cite{Hamada:2014wna}
\aln{
\lambda_\text{eff}(h,h)
	&=	\lambda_\text{min}+{b\ov2}\left(\ln{h\over h_\text{min}}\right)^2.
		\label{lambda approximated}
}
Within this approximation, we see that the potential has a saddle point $V_{}'=V_{}''=0$ at $h=h_s$ when and only when $\lambda_\text{min}$ is tuned to a critical value $\lambda_\text{min}=\lambda_c$ where
\aln{
h_s	&=	e^{-1/4}h_\text{min},\\
\lambda_c
	&=	{b\over32}.
		\label{lambda c given}
}
That is,
\aln{
\lambda_{\rm eff} (h)
	&=	\frac{b}{32}\left[1+16\left(\ln\left(\frac{h}{h_s e^{1/4}}\right)\right)^2\right].
	\label{approximate lambda}
}
We fit $b$ and $h_s$ from the numerical analysis above. 
As a consistency check, we also plot the Higgs potential whose effective coupling is replaced by the analytical one~(\ref{approximate lambda}) in the upper left panel in Fig. \ref{fig:saddle} by dashed black lines.     
One can actually see that the analytical ones well  approximate the effective Higgs potential around  the saddle point. 
In the next section, we will use Eq.~(\ref{lambda approximated}) with $\lambda_\text{min}$ close to $\lambda_c$, as in Eq.~\eqref{Higgs potential in Jordan}, to study the critical Higgs inflation. 
In particular, the minimum value $\lambda_c $ is important because it also determines, along with $h_s $, the value of the potential during the inflation. 
Qualitatively, small $\lambda_c $ will turn out to be favorable to realize successful inflation with small $\xi\lesssim 100$.

In the lower right panel in Fig. \ref{fig:saddle}, we show the numerical calculations of $\lambda_c $ as a function of $\lambda_{SH} $. 
From this plot, one can see that $\lambda_c^{}$ is a monotonically increasing function of $\lambda_{SH} $ and its order of magnitude is $10^{-5}$ around $\lambda_{SH}^{}\sim 0.3$. 
This corresponds to $b={\cal O}(10^{-4})$ by Eq.~(\ref{lambda c given}). 
We will see that the region with small $\lambda_{SH}$ is preferred to realize a successful inflation for small $\xi$.\footnote{
Higgs inflation with $\xi=\mathcal{O}(10^4)$~\cite{Salopek:1988qh,Bezrukov:2007ep} is always possible at the expense of the small cutoff scale $M_\text{P}/\xi$.  
}
Therefore, in the inflationary analysis, we will show the results in the region $\lambda_{SH} \lesssim 0.3$. 
%

\section{Critical Higgs inflation}\label{inflation}
In this section, we study the critical Higgs inflation of our model at the CP 2-2.  
We take full advantage of the saddle point of the Higgs potential. 
The existence of the saddle point makes it easy to obtain the sufficient number of e-foldings $N$ even when $\xi\lesssim 100$;   
contrary to the conventional case, the tensor-to-scalar ratio $r$ does not have to be related to $N$ as $r\sim 1/N^2$, and it can be sizable $\sim 0.05$. 
In section~\ref{results}, we will show our numerical calculations of CMB observables.  

Here we comment on the other fields. 
During the critical Higgs inflation, the Higgs potential is flat, while other scalar fields have large masses due to the couplings  $\lambda_{\phi H}$ and $\lambda_{S H}$ with the large Higgs field, so they do not play a role in the inflation analysis.
As for the renormalization scale, since we are considering the case where only the Higgs field is large, we can consider the Higgs field $h$ as a renormalization point and use the single-scale renormalization group. 
%


\subsection{Higgs inflation at classical level}
We first review the Higgs inflation with non-minimal coupling $\xi h^2 R$ at the classical level~\cite{Salopek:1988qh,Bezrukov:2007ep}.  
We start with the Jordan-frame action
\aln{
S_\text{cl}
	&=	\frac{M_\text{P}^2}{2}\int d^4x\sqrt{-g_J }\Omega^2(h) R_J -\int d^4x\sqrt{-g_J }\left[\frac{1}{2}(\partial h)^2+V_\text{cl}(h) \right],
}
where we have truncated the potential and Weyl factor at the quartic and quadratic orders, respectively:
\aln{
V_\text{cl}(h)
	&=	{\lambda_\text{cl}\over4}h^4,&
\Omega^2(h)
	&=	1+\xi\frac{h^2}{M_\text{P}^2}.
}
By performing the following redefinition of the metric,
\aln{ 
g_{\mu\nu} =\Omega^2g_{J\mu\nu} ,
}
we obtain
\aln{
R_J=\Omega^2
\left[
R
+3\Box\ln \Omega^2 
-{3\over 2}g^{\mu\nu}{(\partial_\mu \ln \Omega^2)(\partial_\nu \ln \Omega^2)}
\right].
}
Then the Einstein-frame action becomes
\aln{
S_\text{cl}
	&=	\int d^4x\sqrt{-g}\left(\frac{1}{2}M_\text{P}^2R
		- \frac{1}{2\Omega^2} 
(\partial h)^2
		-\frac{3}{4} M_\text{P}^2 (\partial_\mu \ln \Omega^2)^2
		-U_\text{cl}(h) + \cdots \right),
\label{eq:action in Einstein frame}
}
where
\aln{
U_\text{cl}(h)
	=	\frac{V_\text{cl}(h)}{\Omega^4(h)}
\label{potential in Einstein frame}
}
is the potential in the Einstein frame.
Since $V_\text{cl}$ is quartic, the following relation holds
\aln{
U_\text{cl}(h)
	=	{\lambda_\tx{cl}\over4}\left(h\over\Omega(h)\right)^4
	=	V_\text{cl}\!\left(h\over\Omega(h)\right).
	\label{classical U and V}
}
In the $h\to\infty$ limit, we have
\aln{
{h\over\Omega(h)}
	&\to	{M_\text{P}\over\sqrt\xi}.
}
Therefore for large $h$, the Einstein-frame potential $U_\tx{cl}(h)$ becomes constant:
\aln{
U_\tx{cl}(h)
	&\to V_\tx{cl}\!\left(M_\tx{P}\over\sqrt\xi\right)
	=	{\lambda_\tx{cl}M_\text{P}^4\over4\xi^2}.
	\label{classical U}
}
This flat potential is used in the Higgs inflation.

The relation between the canonically normalized Einstein-frame field $\chi$ and the Jordan-frame field $h$ is given by
\begin{align}
\frac{d\chi}{dh}=\frac{\sqrt{\Omega^2+6\xi^2 h^2/M_\text{P}^2}}{\Omega^2}.
\end{align}

Under the slow-roll approximation, we obtain
\aln{
\frac{\lambda_\tx{cl} }{\xi^2}\simeq 6.0\times \left(\frac{50}{N}\right)^2\times 10^{-10}
}
to fit $A_s=U/(24\pi^2M_\text{P}^4\varepsilon_V)$ to the observed value $2.1\times10^{-9}$ at the $e$-folding $N$; see~\ref{ordinary Higgs inflation results}.
We see that the typical SM value at low energy $\lambda_\text{cl}\sim 0.1$ requires large value of $\xi\sim 10^5$.

\subsection{Higgs inflation including radiative correction}\label{Higgs inflation with radiative}
At the quantum level in the flat spacetime, we promote $V_\tx{cl}(h)$ to the effective potential
\aln{
V(h)=V_\text{tree}(h,\mu)+\Delta V_\text{loop}(h,\mu),
	\label{flat space effective potential}
}
where $V_\text{tree}$ is the tree-level potential including the field renormalization and $\Delta V_\text{loop}$ is the loop correction; see Eq.~(\ref{full 1-loop potential}) for the 1-loop approximation.
In this paper, we employ the Einstein-frame effective potential on the so-called Prescription I,
\aln{
U(h)
	&=
	{\lambda_H (\mu)e^{4\Gamma(\mu)}h^4\over4\Omega^4(h)}
	+\Delta U_\text{loop}\!\left({h},\mu\right),
}
where $\Delta U_\text{loop}$ is obtained from $\Delta V_\text{loop}$ in Eq.~\eqref{flat space effective potential} by replacing all the effective masses $M_\Psi(h)$ with ${M_\Psi(h)\over\Omega}$ for $\Psi=W$,$Z$, $t$, $\nu i$, $S$, and $\phi$:
\aln{
\Delta U_\text{loop}\!\left({h},\mu\right)
	&=	\Delta V_\text{loop}\!\left(h,\mu\right)\Big|_{M_\Psi(h)\to{M_\Psi(h)\over\Omega}}.
}
See Ref.~\cite{Hamada:2016onh} for the meaning of this prescription as well as ambiguity due to indeterminacy of non-renormalizable terms. Because the potential should not depend on $\mu$ (if one takes all order corrections into account), we may replace $\mu$ by $\mu/\Omega$: 
\footnote{In the second line of Eq.~(\ref{potential in prescription I}), we have assumed that the $\Omega$ dependence in $\Delta V_{\rm loop}^{}(h,\mu/\Omega)$ disappears by the replacement $M_\psi^{}(h)\rightarrow M_\psi^{}(h)/\Omega$. 
At one loop level, we can easily check this from Eq.~(\ref{one-loop Higgs potential}). 
}
\aln{
U_{}(h)
	&=
	{\lambda_H ({\mu\over\Omega})e^{4\Gamma({\mu\over\Omega})}h^4\over4\Omega^4}
	+\left.\Delta V_\text{loop}\!\left({h},{\mu\over\Omega}\right)\right|_{M_\Psi(h)\to{M_\Psi(h)\over\Omega}}\nn
	&=
	{\lambda_H ({\mu\over\Omega})e^{4\Gamma({\mu\over\Omega})}h^4\over4\Omega^4}
	+{1\ov\Omega^4}\Delta V_\text{loop}\!\left({h},\mu\right).
\label{potential in prescription I}
}
When we truncate the loop correction at the 1-loop order, we can obtain better approximation by choosing $\mu$ to be around $h$ to minimize the higher loop corrections:
\aln{
U_{}(h)
	&=
	{1\ov\Omega^4}\left({\lambda_H \!\left(h\over\Omega\right)\over4}e^{4\Gamma\!\left(h\over\Omega\right)}h^4
	+\Delta V_\text{1-loop}\!\left(h,h\right)\right).
}
This is the expression we employ in the following.
When $y_{\nu i}h$ is sufficiently larger than $M_R$, all the effective masses are proportional to $h$, therefore $\Delta V_\text{1-loop}(h,h)$ is proportional to $h^4$ whose coefficient is independent of $h$ (up to higher order corrections), and hence we obtain 
\aln{
U_{}(h)
	&=	V_{} \!\left({h\over\Omega}\right).
	\label{relation in prescription I}
}
Thus, after taking into account the quantum corrections, the same relation still holds as Eq.~\eqref{classical U and V} on Prescription I, and again the Einstein-frame potential~\eqref{relation in prescription I} becomes constant for $\xi h^2/M_\text{P}^2\gg1$:
\aln{
U(h)
	&=	{\lambda_\text{eff}\!\left({h\over\Omega(h)},{h\over\Omega(h)}\right)\over4}\left(h\over\Omega(h)\right)^4
	\to	{\lambda_\text{eff}\!\left({M_\text{P}\over\sqrt\xi},{M_\text{P}\over\sqrt\xi}\right)\over4}\left(M_\text{P}\over\sqrt\xi\right)^4.
}
Unless $\lambda_\text{eff}\!\left({M_\text{P}\over\sqrt\xi},{M_\text{P}\over\sqrt\xi}\right)$ is particularly small, we still need large $\xi$ to fit $A_s$.
In fact in the SM, it is known that $\lambda_\text{eff}(h,h)$ becomes small around $h\sim M_\text{P}$, which is an essential ingredient in the critical Higgs inflation.

\subsection{Critical Higgs inflation}
As can be seen from Eq.~\eqref{relation in prescription I}, when $V_{} (h)$ has a saddle point at $h=h_s $, $U_{}(h)$ also has a saddle point at $h=\tilde h_s$, with $\tilde h_s$ being determined by ${\tilde h_s\over\Omega(\tilde h_s)}=h_s$:
\aln{
\tilde{h}_s=\frac{h_s }{1-c_s^2},
}
where we have introduced
\aln{
c_s:=h_s{\sqrt\xi\over M_\text{P}}. 
	\label{cs defined}
}
This parameter is the ratio of $h_s $ to $M_\text{P}/\sqrt\xi$; the latter is the typical value of $h$ above which the conformal factor $\Omega(h)$ starts to deviate from unity.   
One can see that $\tilde{h}_s $ approaches infinity as $c_s\nearrow 1$, which means that the small region around the saddle point $h=h_s $ is widely stretched, and allows a sufficient $e$-folding.

In the critical Higgs inflation, we assume that the high-scale Higgs potential is close to a one having a saddle point, namely, $\lambda_\text{min}$ in Eq.~\eqref{lambda approximated} is close to $\lambda_c$ given by Eq.~\eqref{lambda c given}:
\aln{
\lambda_{\rm min}
	&=	\left(1+\delta\right)\lambda_c,
\label{small parameter}
}
where we have parametrized the deviation from the saddle-point criticality by $\delta$.
Then the flat-space effective potential becomes
\aln{
V_{}(h) =\frac{\lambda_{\rm eff} (h,h)}{4}h^4=\frac{\lambda_c }{4}\left[1+\delta +16\left(\ln\left(\frac{h}{h_s e^{1/4}}\right)\right)^2\right]h^4.
\label{Higgs potential in Jordan}
}

Using Eqs.~\eqref{relation in prescription I} and \eqref{cs defined}, we see that $U_{}$ approaches the constant value in the $h\to\infty$ limit
\aln{
U_{}(h)
	&\to
		\frac{\lambda_c }{4}\left[1+\delta +\left(1+4\ln c_s\right)^2\right]{M_\text{P}^4\over\xi^2},
		\label{Vinf}
}
which determines the value of the potential during the inflation. 
In this paper, we will focus on $c_s\leq 1$ and take full advantage of the saddle point of the Higgs potential. 

\begin{figure}[t!]
\begin{center}
\includegraphics[width=10cm]{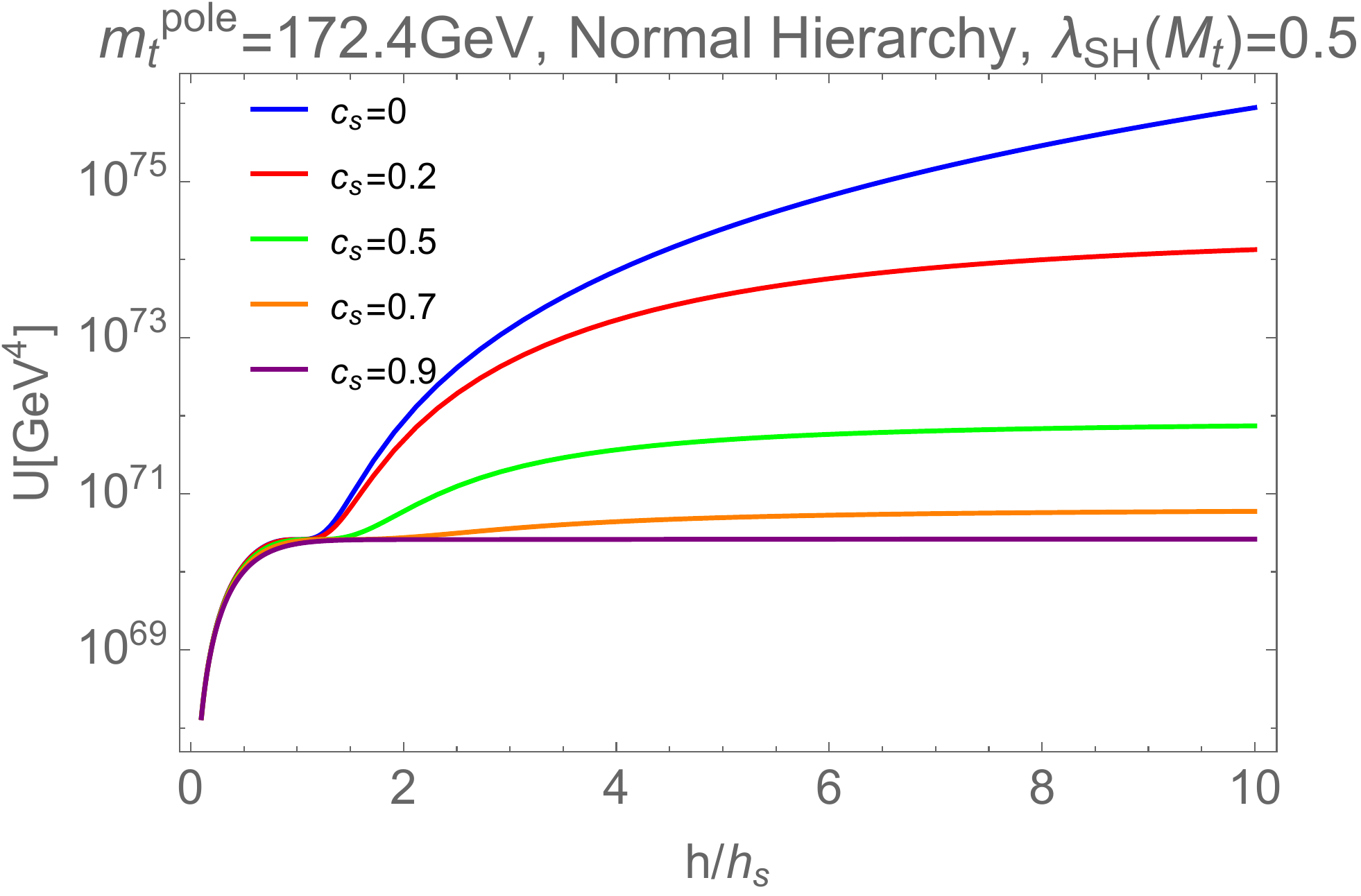}
\caption{
The Higgs potential in the Einstein frame. 
}
\label{fig:potential_Einstein}
\end{center}
\end{figure}

In Fig.~\ref{fig:potential_Einstein}, we show the Higgs potential in the Einstein frame where different colors correspond to different values of $c_s$. 
Here, we show the the normal hierarchy case with $\lambda_{SH} (M_t )=0.5$ for illustration.  
One can see that the region around the saddle point $h=h_s $ is more and more stretched as we increase $c_s$ toward unity. 
Therefore, for a given $c_s\lesssim 1$, we may always fit the $e$-folding $N$ around $h=h_s\Omega$ by tuning $\delta$.\footnote{
The tuning of $\lambda_\text{min}$ to $\lambda_c$ is favored by the maximum entropy principle because the more the space is expanded by the inflation, the more the total entropy emerges~\cite{Kawai:2011qb,Hamada:2015dja}.
}

\section{Prediction on inflationary observables}\label{Prediction on inflationary observables}
Here, we analyze the prediction on inflationary observables as we vary the parameters in the model. So far, we have three free parameters: $\lambda_{SH}$, $M_R$, and $\xi$. Recall that the scalar sector has only one parameter $\lambda_{SH}$ in our analysis on the red dotted line in Fig.~\ref{fig:CP22}.

On the other hand, we have seen that the Jordan-frame potential can be parametrized near the saddle-point criticality by $\lambda_c$, $\delta$, and $h_s$ as Eq.~\eqref{Higgs potential in Jordan}. Here, these three parameters are functions of the model parameters $\lambda_{SH}$ and $M_R$.

In order to sweep the parameters near the saddle point, we use the results of Sec.~\ref{sec:saddle}: For each $\lambda_{SH}$, we find the value of $M_R$ that gives the saddle-point criticality, $\delta=0$, as well as the corresponding parameters $h_s$ and $\lambda_c$.
See the upper-right, lower-left, and lower-right panels in Fig.~\ref{fig:saddle} for $M_R$, $h_s$, and $\lambda_c$, respectively. 
For a given $\lambda_{SH}$, as we slightly change $M_R$ from the critical value, in general all of the parameters $\lambda_c$, $h_s$, and $\delta$ are modified. Here, we neglect the change of $\lambda_c$ and $h_s$, and take into account the effect of non-zero $\delta$.

Now we take into account the non-minimal coupling $\xi$.
Among three parameters $\xi, \lambda_{SH}, \delta$, the last one is traded with $e$-folding $N$ using Eq.~\eqref{e-folding}.
The observables $n_s$ and $r$ are functions of $\xi$ and $\lambda_{SH}$ once $e$-folding $N$ is fixed by $\delta$.

\subsection{Results} \label{results}
\begin{figure}[t!]
\begin{center}
\includegraphics[width=0.48\textwidth]{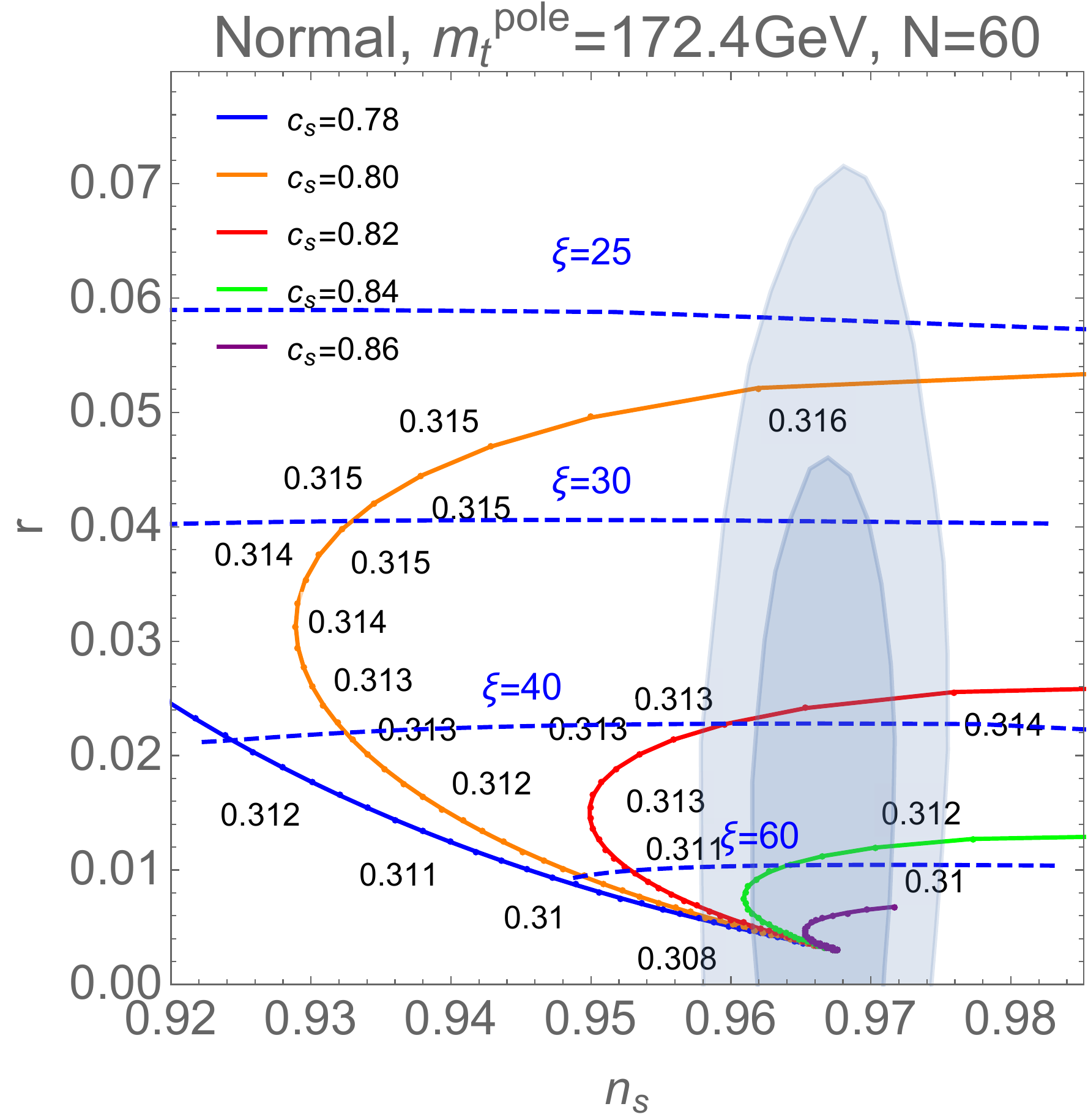}
\includegraphics[width=0.48\textwidth]{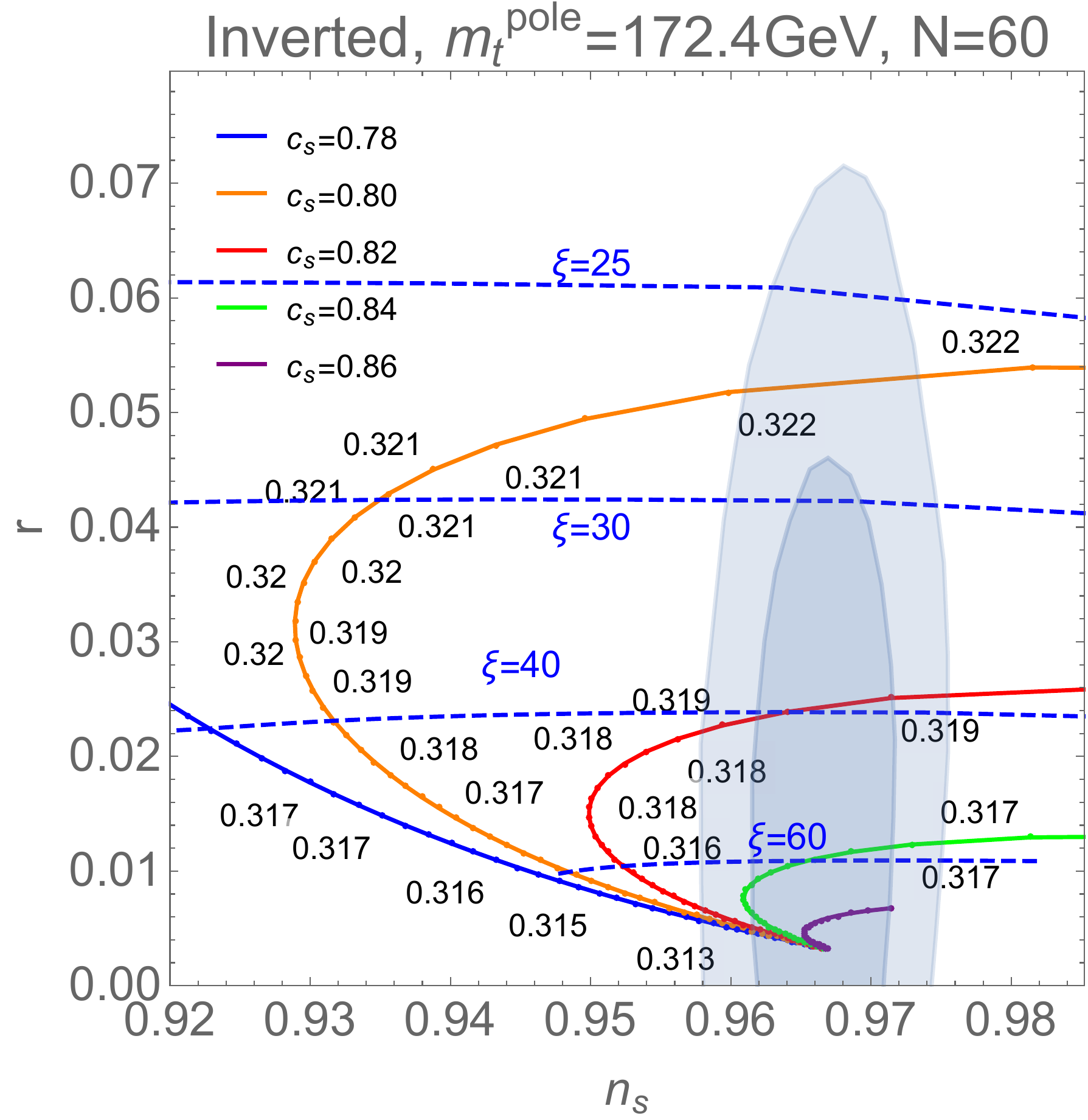}
\\
\includegraphics[width=0.48\textwidth]{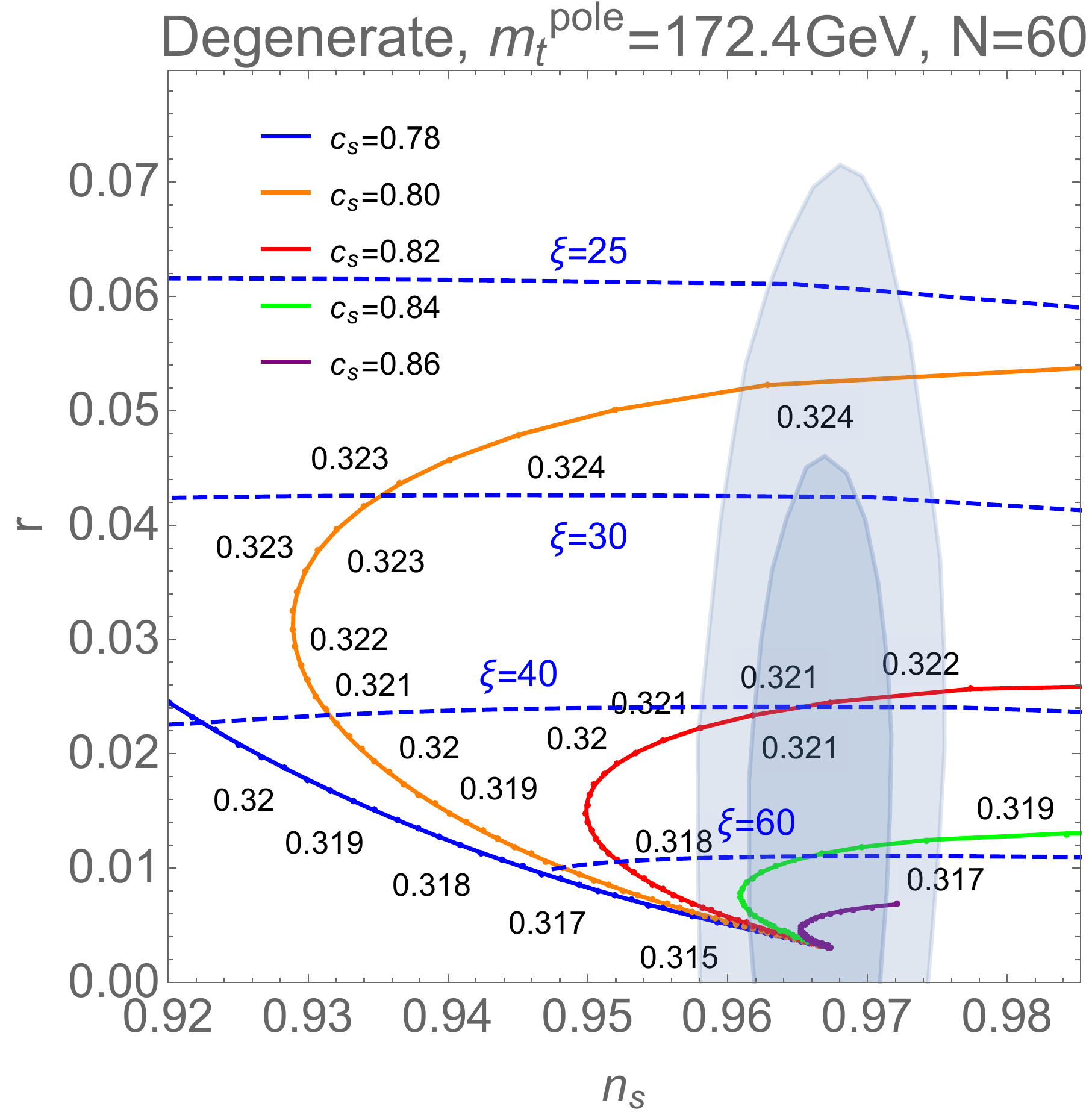}
\caption{Upper Left (Right): $r$ vs $n_s $ in the case of the normal (inverted) hierarchy where $c\ (\xi)$ is fixed on each solid (dashed) curves and $\lambda_{SH} (M_t )$ is varied. 
The blue regions correspond to the allowed regions by Planck 2018. 
 \\
 Lower: the degenerate case. 
}
\label{fig:ns-r}
\end{center}
\end{figure}
In Fig. \ref{fig:ns-r}, 
we show the values of $\xi$, $c_s$, and $\lambda_{SH}$ in the  $n_s$-$r$ plane for $N=60$ with the central value $m_t^\text{pole}=172.4$ GeV: The upper left (right) panel corresponds to 
the normal (inverted) hierarchy case, and the lower panel to the degenerate case.  
The values of $\xi$ and $c_s$ are shown by the solid and dashed lines, respectively, while $\lambda_{SH}$ by the numbers on the solid line.
The dark (light) blue region is allowed by the combined analysis of Planck 2018 at the $65\%\ (95\%)$ CL.
From these results, one can see that our model at the CP 2-2 is consistent with the current CMB observations even when $\xi=25$.  
The smaller the $r$, the larger the required value of $\xi$: If the upper bound becomes $r<0.04$ (0.02), we need $\xi\gtrsim 30$ (40).  

We see that we typically have $\lambda_{SH} \sim 0.32$, which corresponds to the large Majorana mass as
\aln{
5\times 10^{13}{\rm GeV}\lesssim M_R \lesssim 2\times 10^{14}{\rm GeV}
}
from the upper right panel in Fig. \ref{fig:saddle}.  

\begin{figure}[t!]
\begin{center}
\includegraphics[width=0.48\textwidth]{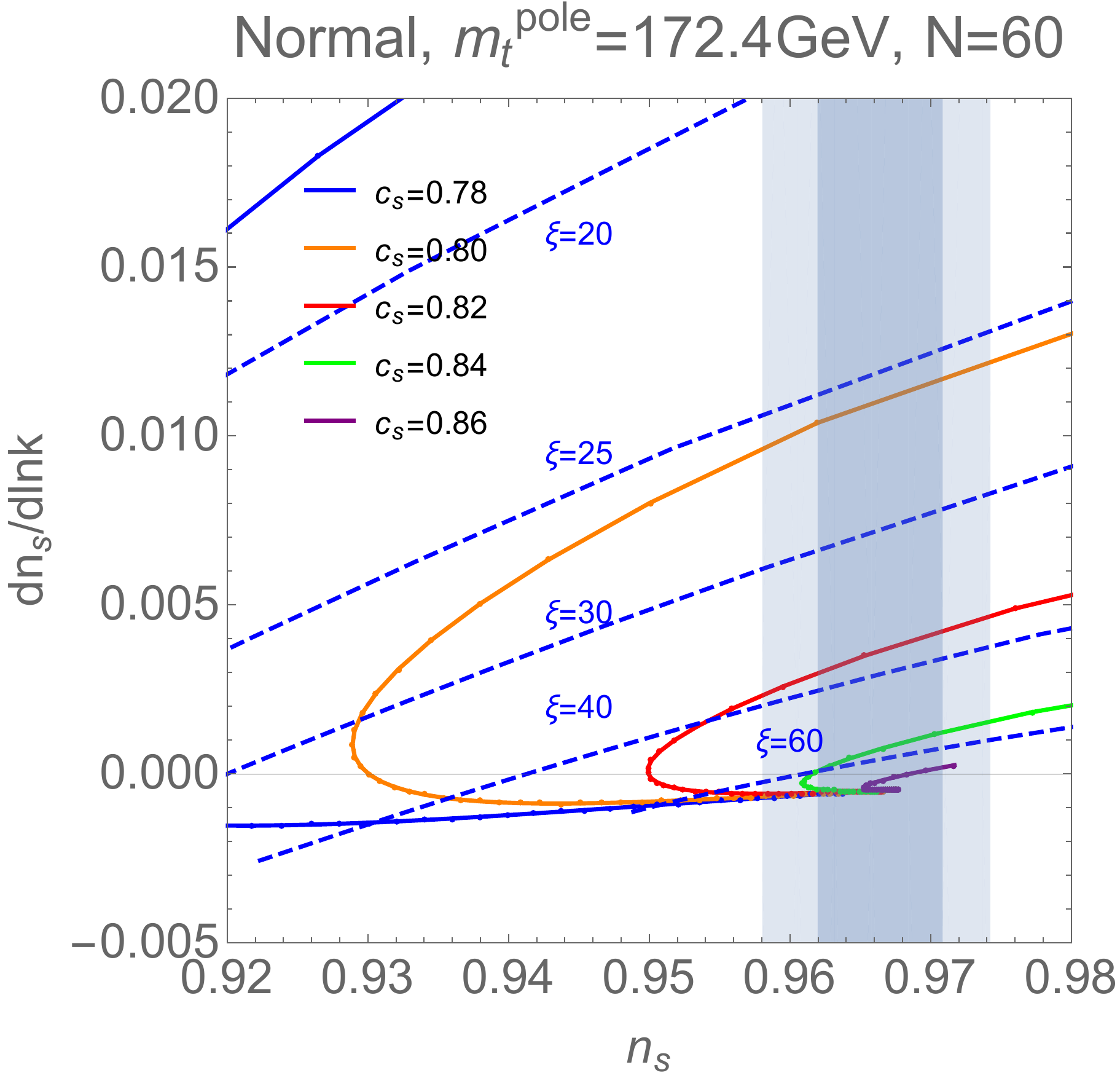}
\includegraphics[width=0.48\textwidth]{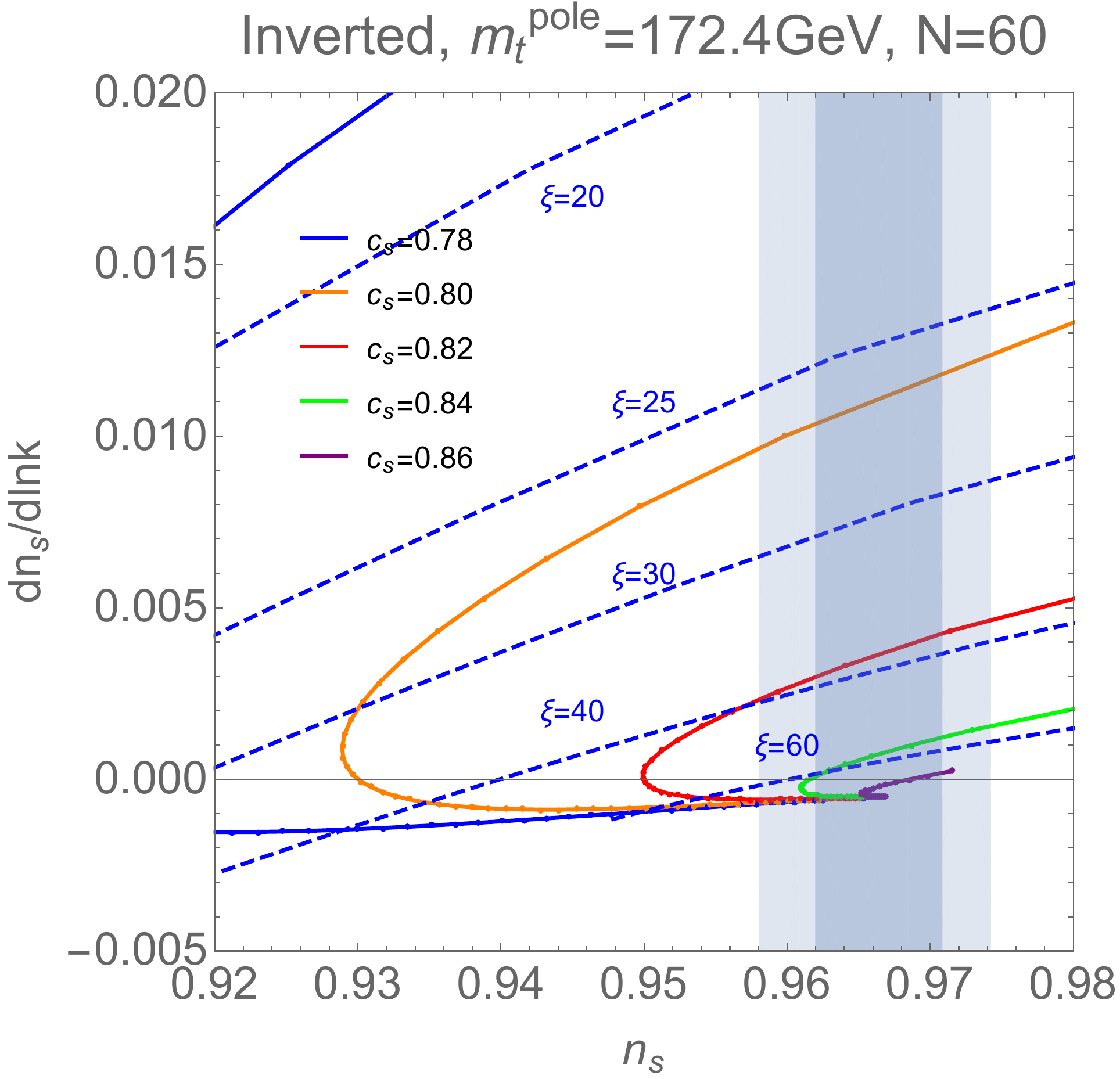}
\\
\includegraphics[width=0.48\textwidth]{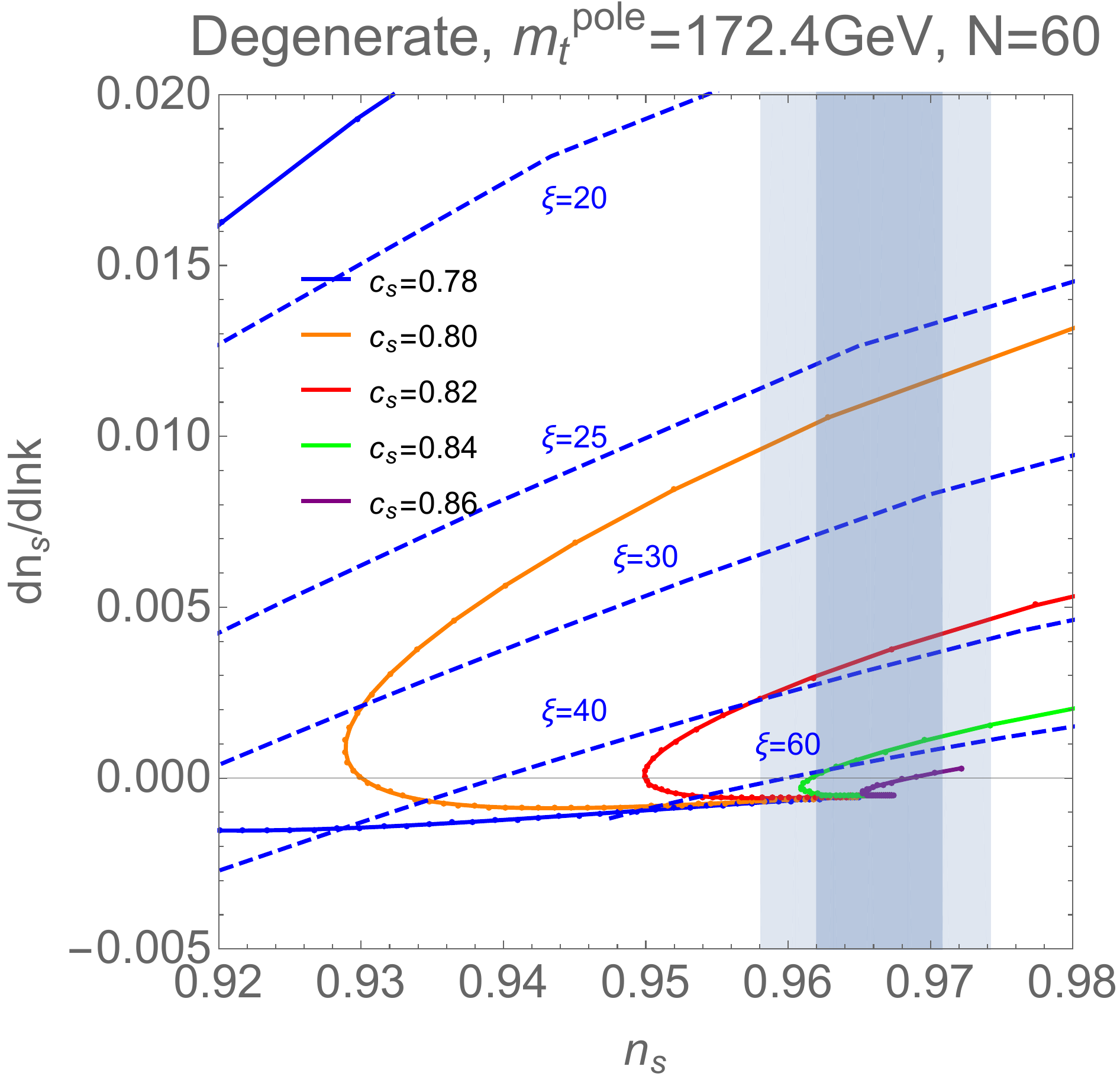}
\caption{Upper Left (Right): $n_s $ vs $dn_s /d\ln k$ in the case of the normal (inverted) hierarchy where $c\ (\xi)$ is fixed on each solid (dashed) curves. 
The blue regions correspond to the allowed regions by Planck 2018. 
 Lower: the degenerate case. 
}
\label{fig:ns-dns}
\end{center}
\end{figure}

In Fig.\ref{fig:ns-dns}, we also show $dn_s /d\ln k$ vs $n_s $, where $c_s$ and $\xi$ are again shown as in Fig.~\ref{fig:ns-r}. 
In this case, the observational error of $dn_s /d\ln k$ is still too large to constrain the inflaton potential.   

\newpage
\section{Summary and discussion}\label{summary}
Motivated by various fundamental issues in particle physics and cosmology, we have discussed the minimal model that can explain EW scale, neutrino masses, DM, and successful inflation at the same time. 
The model adds right-handed neutrinos to the two-scalar model in Refs.~\cite{Haruna:2019zeu,Hamada:2020wjh}, which has been proposed to explain the origin of EW scale and DM. 
These two scalar fields give a minimal setup to realize an analogue of the CW mechanism.
Assuming the $Z_2 $ symmetry of a scalar, $S\to-S$, it can be a candidate for DM, similarly to the Higgs-portal scalar DM model. 
Neutrino masses are naturally explained by the seesaw mechanism.

In this paper, we have analyzed RGEs, calculated the effective Higgs potential, and studied the critical Higgs inflation that uses the (near) saddle point of the Higgs potential at a high scale. 
The new scalar coupling $\lambda_{SH} $ between the SM Higgs and DM $S$ can stabilize the Higgs potential even if the top mass is current center value $m_t =172.4$ GeV. 
In our model, it is possible to maintain the existence of saddle point by the neutrino Yukawa coupling $y_\nu $, and the saddle point condition relates the parameters of the model as is shown in Fig. \ref{fig:saddle}. 

By utilizing the saddle point of the Higgs potential, we have found that it is possible to realize successful inflation even for $\xi\sim 25$ within the parameter space where all the necessary requirements are satisfied.   
As a result, we obtain $\lambda_{SH}\simeq 0.32$ and $\langle\phi\rangle\simeq 2.7$\,TeV, which correspondingly lead to the dark matter mass $m_S \sim 2.0$\,TeV, its spin-independent cross section $1.8\times10^{-9}$\,pb, and the mass of additional neutral scalar $m_H \sim 190$\,GeV.  

Finally, we mention testability of our model at collider experiments and future directions of this scenario.   
Since the DM should be as heavy as a TeV range in order to satisfy the relic abundance and the constraints from the direct searches, the detection of the extra Higgs can be an important probe of our model similarly to the Higgs singlet model. 
On the benchmark points shown in Fig. 1, the mass of the additional Higgs boson is predicted to be in the range of 70 GeV -- 200 GeV, so that it can be produced at future lepton colliders such as the International Linear Collider (ILC) via the Z boson strahlung process. 
Therefore, our model can be tested at the ILC and/or its energy upgraded version. 

Our model can be well tested at the near future DM detection experiments such as XENONnT. 
If the whole region is excluded by them, one of the simplest extensions would be introduction of extra heavy fermions that are singlet under the SM gauge symmetry but are odd under $Z_2$ such that they lower the quartic scalar couplings in the RGE running to make the perturbativity bound milder.

It would be interesting to analyze all the possible criticalities along the line of the current work. 
Moreover, we can also come up with a lot of interesting phenomena within this model such as a possibility of producing primordial Black Hole by Higgs inflation \cite{Ezquiaga:2017fvi,Bezrukov:2017dyv,Rasanen:2018fom,Cheong:2019vzl}, spontaneous leptogenesis \cite{Lee:2020yaj,Kusenko:2014lra,Cohen:1987vi,Cohen:1988kt,Dolgov:1994zq}, (p)reheating dynamics and so on. 
We would like to discuss those possibilities in future investigations. 

Finally, we comment on possible systematic errors introduced by higher-order corrections. We have used the one-loop effective potential to discuss the MPP at the TeV-scale, and two-loop RGEs to extrapolate from there to high scales. The higher-order corrections are further suppressed at approximately $\lambda_{SH}/8\pi^2\lesssim$ few percent, compared to the corrections currently considered.  The same degree of corrections applies to the potential value $\propto \lambda_H^{}$ and the inflationary predictions derived from it. 

\section*{Acknowledgements} 
The work of YH is supported by JSPS Overseas Research Fellowships.
The work of KO is partly supported by the JSPS Kakenhi Grant No.~19H01899.
The work of KY is supported in part by the Grant-in-Aid for Early-Career Scientists, No.~19K14714. 
%

\appendix 
\def\thesection{Appendix \Alph{section}}
\section{Two-Loop RGEs}\label{RGEs}
Here, we summarize the two-loop RGEs.  
Our calculations are based on \cite{Machacek:1983tz,Machacek:1983fi,Machacek:1984zw,Luo:2002ti}. 
%
\aln{
\frac{dg_Y }{dt}&=\frac{1}{(4\pi)^2}\frac{41}{6}g_Y^3+\frac{g_Y^3}{(4\pi)^4}\frac{19}{6}\left(\frac{199}{18}g_Y^2+\frac{9}{2}g_2^2+\frac{44}{3}g_3^2-\frac{17}{6}y_t^2-\frac{n_\nu }{2}y_\nu^2\right),
\\
\frac{dg_2 }{dt}&=\frac{1}{(4\pi)^2}\left(-\frac{19}{6}\right)g_2^3+\frac{g_2^3}{(4\pi)^4}\frac{19}{6}\left(
\frac{3}{2}g_Y^2+\frac{35}{6}g_2^2+12g_3^2-\frac{3}{2}y_t^2-\frac{n_\nu }{2}y_\nu^2\right),
\\
\frac{dg_3 }{dt}&=\frac{1}{(4\pi)^2}\left(-7\right)g_2^3+\frac{g_2^3}{(4\pi)^4}\frac{19}{6}\left(
\frac{11}{6}g_Y^2+\frac{9}{2}g_2^2-26g_3^2-2y_t^2\right),
\\
\frac{d\lambda_H }{dt}&=\frac{1}{16\pi^2}\left(
24\lambda_H^2+
\frac{\lambda_{SH}^2}{2}+\frac{\lambda_{\phi H}^2}{2}-3\lambda_H (g_Y^2+3g_2^2)+\frac{3}{8}g_Y^4+\frac{3}{4}g_Y^2g_2^2+\frac{9}{8}g_2^4+4\lambda_H (3y_t^{2}+n_\nu y_\nu^{2})-6y_t^4-2n_\nu y_\nu^4
\right)
\nn
&+\frac{1}{(4\pi)^4}\bigg\{-312 \lambda_H^3-2 \lambda _{\phi H}^3-5
   \lambda_H  \lambda_{\phi H}^2-2 \lambda_{SH}^3-5 \lambda_H \lambda
   _{SH}^2+36 \lambda_H^2 \left(g_Y^2+3 g_2^2\right)
   \nn
   &\quad \quad -\lambda_H \left(-\frac{39}{4} g_2^2 g_Y^2-\frac{629 g_Y^4}{24}+\frac{73 g_2^4}{8}\right)
  +\frac{305
   g_2^6}{16} -\frac{289}{48} g_2^4 g_Y^2-\frac{559}{48} g_2^2 g_Y^4-\frac{379 g_Y^6}{48}
  -32 g_3^2 y_t^4 -\frac{8}{3} g_Y^2 y_t^4
  \nn
  &\quad\quad -\frac{9}{4} g_2^4 (y_t^2+y_{\nu}^2)+g_Y^2y_t^2 \left(\frac{21 g_2^2}{2}-\frac{19 g_Y^2}{4}\right)+n_\nu g_Y^2y_{\nu }^2 \left(-\frac{g_Y^2}{4}-\frac{ g_2^2}{2}\right) +\lambda_H  y_t^2 \left(\frac{85 g_Y^2}{6}+\frac{45 g_2^2}{2}+80 g_3^2\right) 
  \nn
  &\quad\quad +n_\nu \lambda_H  y_{\nu }^2 \left(\frac{5 g_Y^2}{2}+\frac{15 g_2^2}{2}\right) - 48\lambda_H^2 \left(3y_t^2+n_\nu y_{\nu }^2\right)-\lambda_H \left(3y_t^4+n_\nu y_{\nu}^4\right)+10
   \left(3y_t^6+n_\nu y_{\nu }^6\right)
\bigg\},\label{lambda H beta function}
\\
\frac{d\lambda_\phi}{dt}&=\frac{1}{16\pi^2}\left(3\lambda_\phi^2+3\lambda_{\phi S}^2+12\lambda_{\phi H}^2
\right)
+\frac{1}{(4\pi)^4}\bigg\{
-\frac{17 \lambda _{\phi }^3}{3}-20 \lambda
   _{\phi } \lambda _{\phi H}^2-48 \lambda _{\phi H}^3-12 \lambda
   _{\phi S}^3 -5 \lambda _{\phi } \lambda _{\phi S}^2
      \nn
&\quad \quad 
-72 \lambda _{\phi H}^2 y_t^2-24n_\nu \lambda _{\phi H}^2 y_{\nu}^2
+72 g_2^2 \lambda _{\phi H}^2+24 g_Y^2 \lambda _{\phi H}^2
\bigg\},
\\
\frac{d\lambda_S }{dt}&=\frac{1}{16\pi^2}\left(3\lambda_S^2+3\lambda_{\phi S}^2+12\lambda_{S H}^2
\right)
+\frac{1}{(4\pi)^4}\bigg\{-\frac{17 \lambda _S^3}{3}-20 \lambda
   _S \lambda _{SH}^2-48 \lambda _{SH}^3-12 \lambda _{\text{$\phi
   $S}}^3
-5 \lambda _{\phi S}^2 \lambda_S
   \nn
&\quad \quad
 -72 \lambda _{SH}^2 y_t^2-24n_\nu  
   \lambda _{SH}^2 y_{\nu }^2+72 g_2^2 \lambda _{SH}^2+24 g_Y^2 \lambda _{SH}^2
\bigg\},
}
\aln{
\frac{d\lambda_{\phi H} }{dt}&=\frac{\lambda_{\phi H} }{16\pi^2}\left(12\lambda_H +\lambda_{\phi} -4\lambda_{\phi H} +6y_t^2
+2n_\nu y_\nu^2
-\frac{3}{2}g_Y^2-\frac{9}{2}g_2^2\right)+\frac{\lambda_{SH} \lambda_{\phi S} }{16\pi^2}
\nn
&\quad +\frac{\lambda _{\phi H}}{(4\pi)^4}\bigg\{-\frac{21 \lambda _{\phi H}^2}{2}
-72 \lambda_H   \lambda _{\phi H}-60 \lambda_H^2 -6 \lambda _{\phi} \lambda _{\phi H}-\frac{5}{6} \lambda _{\phi }^2 
   -\frac{1}{2} 
   \lambda _{\phi S}^2-4 
   \lambda _{\phi S }\lambda _{SH}-\frac{1}{2} 
   \lambda _{SH}^2
   \nn
  &\quad\quad
-\left(4 \lambda _{\phi H}+24 \lambda_H   
   \right)
   (3y_t^2+n_\nu y_{\nu }^2)
-\frac{9}{2} 
   (3y_t^4+n_\nu  y_{\nu }^4) +g_Y^2 
  \left(\lambda _{\phi H}+24 \lambda_H 
   \right)
   +3 g_2^2 
   \left(\lambda _{\phi H}+24 \lambda_H 
   \right)
   \nn
&\quad\quad +
   y_t^2 \left(\frac{85 g_Y^2}{12}+\frac{45 g_2^2}{4}+40
   g_3^2\right)
+n_\nu y_{\nu }^2\left(\frac{5 g_Y^2}{4}+\frac{15 g_2^2}{4}\right) 
-\frac{145}{16} g_2^4 
+\frac{15}{8} g_2^2 g_Y^2 
+\frac{557}{48} g_Y^4 
\bigg\}
\nn
&\quad\quad+\frac{1}{(4\pi)^4}\left(-2 \lambda _{\phi S} \lambda _{SH}^2-2 \lambda _{\phi S}^2
   \lambda _{SH}\right),
\\
\frac{d\lambda_{SH}}{dt}&=\frac{\lambda_{S H} }{16\pi^2}\left(12\lambda_H +\lambda_{S} +4\lambda_{S H} +6y_t^2
+2n_\nu y_\nu^2
-\frac{3}{2}g_Y^2-\frac{9}{2}g_2^2
\right)-\frac{\lambda_{\phi H} \lambda_{\phi S} }{16\pi^2}
\nn
&\quad 
+\frac{\lambda _{SH}}{(4\pi)^4}\bigg\{-\frac{21 \lambda _{SH}^2}{2}
-72 \lambda_H   \lambda _{SH}-60 \lambda_H^2 -6 \lambda _{S}  \lambda _{SH}-\frac{5}{6} \lambda _{S}^2 
   -\frac{1}{2} 
   \lambda _{\phi S}^2-4 
   \lambda _{\phi S} \lambda _{\text{$\phi$H}}-\frac{1}{2} 
   \lambda _{\text{$\phi$H}}^2
   \nn
  &\quad 
-\left(4 \lambda _{SH}+24 \lambda_H   
   \right)
   (3y_t^2+n_\nu y_{\nu }^2)
-\frac{9}{2} 
   (3y_t^4+n_\nu y_{\nu }^4)+g_Y^2 
  \left(\lambda _{SH}+24 \lambda_H 
   \right)
   +3 g_2^2 
   \left(\lambda _{SH}+24 \lambda_H 
   \right)   
\nn
&\quad\quad +
   y_t^2 \left(\frac{85 g_Y^2}{12}+\frac{45 g_2^2}{4}+40
   g_3^2\right)+n_\nu y_{\nu }^2\left(\frac{5 g_Y^2}{4}+\frac{15 g_2^2}{4}\right) 
 -\frac{145}{16} g_2^4 
+\frac{15}{8} g_2^2 g_Y^2 
+\frac{557}{48} g_Y^4 
\bigg\}
\nn
&\quad \quad+\frac{1}{(4\pi)^4}\left(-2 \lambda _{\phi H}^2 \lambda _{\phi S}-2 \lambda _{\phi H}
   \lambda _{\phi S}^2\right),
\\
\frac{d\lambda_{\phi S}}{dt}&=\frac{\lambda_{\phi S} }{16\pi^2}\left(\lambda_{\phi}+\lambda_{S} +4\lambda_{\phi S} 
\right)-\frac{\lambda_{\phi H}^{}\lambda_{SH}^{}}{16\pi^2}
+\frac{\lambda _{\phi S}}{(4\pi)^4}\bigg\{-9 \lambda _{\phi S}^2
-6 \lambda_{\phi S} \left(\lambda _{\phi }+\lambda _S\right)-\frac{5}{6} \left(\lambda _{\phi }^2+\lambda _S^2\right)
\nn
&-2 \left(\lambda _{\phi H}^2+\lambda _{SH}^2\right)-16 \lambda _{\phi H} \lambda _{SH}
\bigg\}+\frac{\lambda _{\phi H} \lambda _{SH}}{(4\pi)^4}\left\{-8\left(3y_t^2+n_\nu y_{\nu }^2\right)+8(g_Y^2+3g_2^2)-8(\lambda_{SH} +\lambda_{\phi H} )
\right\},
}
\aln{
\frac{dy_t }{dt}&=\frac{y_t }{16\pi^2}\left(\frac{9}{2}y_t^2+n_\nu y_\nu^2-\frac{17}{12}g_y^2-\frac{9}{4}g_2^2-8g_3^2\right)
+\frac{y_t }{(4\pi)^4}\bigg\{-12 y_t^4-\frac{9n_\nu }{4} y_t^2 y_{\nu }^2-\frac{9n_\nu }{4}y_{\nu}^4+6
   \lambda_H^2+\frac{\lambda _{\phi H}^2}{4}+\frac{\lambda _{SH}^2}{4}
   \nn
  &\quad\quad -12
   \lambda_H   y_t^2
   +\frac{131}{16} g_Y^2 y_t^2+\frac{225}{16} g_2^2 y_t^2+36 g_3^2 y_t^2
   +n_\nu \left(\frac{15}{8} g_2^2
  +\frac{5}{8} g_Y^2\right) y_{\nu }^2
   \nn
&\quad\quad  +\frac{1187
   g_Y^4}{216}-\frac{23 g_2^4}{4} -108 g_3^4-\frac{3}{4} g_2^2 g_Y^2+9 g_3^2 g_2^2+\frac{19}{9} g_3^2 g_Y^2
\bigg\},
\\
\frac{dy_\nu }{dt}&=\frac{y_\nu }{16\pi^2}\left(\left(n_\nu +\frac{3}{2}\right)y_\nu^2+3y_t^2
-\frac{3}{4}g_y^2-\frac{9}{4}g_2^2\right)
+\frac{y_\nu }{(4\pi)^4}\bigg\{-\left(\frac{9n_\nu }{2}-\frac{3}{2}\right) y_{\nu }^4-\frac{27}{4} y_t^2 y_{\nu }^2-\frac{27
   }{4}y_t^4
   \nn
   &\quad\quad
   +6 \lambda_H^2+\frac{\lambda _{\phi H}^2}{4}
   +\frac{\lambda_{SH}^2}{4}-12 \lambda_H   y_{\nu }^2
   +g_Y^2\left(\frac{85}{24}y_t^2+\left(\frac{5n_\nu }{8}+\frac{93}{16}\right) y_{\nu }^2\right)+g_2^2\left(\frac{45}{8}  y_t^2+\left(\frac{15n_\nu }{8}+\frac{135}{16}\right)
   y_{\nu }^2\right)
   \nn
   &\quad\quad+20 g_3^2 y_t^2+\frac{35
   g_Y^4}{24}-\frac{23 g_2^4}{4}-\frac{9}{4} g_2^2 g_Y^2
\bigg\},
\\
\gamma_H &=\frac{y_\nu }{16\pi^2}\left(\frac{9}{4}g_2^2+\frac{3}{4}g_Y^2-3y_t^2-n_\nu y_\nu^2\right).
\label{gammaH}
}

\section{Single-field slow-roll inflation}\label{Single-field slow-roll inflation}
Here we summarize basic results for the single-field slow-roll inflation.

The slow roll parameters are defined by 
\aln{
\varepsilon_V 
	&=	\frac{M_\text{P}^2}{2}\left(\frac{U_{}^{\prime}}{U_{}}\right)^2,&
\eta_V 
	&=	\frac{M_\text{P}^2}{2}\frac{U_{}^{\prime\prime}}{U_{}},&
\zeta_V^{2}
	&=	M_\text{P}^4\left(\frac{U_{}^{\prime\prime\prime}}{U_{}}\right)\left(\frac{U_{}^{\prime}}{U_{}}\right),
	\label{slow roll parameters}
}
where $U_{}$ is the inflation potential in the Einstein frame and the prime represents the derivative with respect to $\chi$. 

The number of e-foldings from a field value $\chi_*$ to the end of inflation is given by
\aln{
N=\int dt H\simeq \frac{1}{M_\text{P} }\int_{\chi_{\rm end} }^{\chi_*}\frac{d\chi}{\sqrt{2\varepsilon_V}}. 
\label{e-folding} 
}
The CMB observables are given by
\aln{
A_s 
	&=	\frac{U_{}}{24\pi^2M_\text{P}^4\varepsilon_V },&
r	&=	16\varepsilon_V ,&
n_s 
	&=	1-6\varepsilon_V +2\eta_V ,&
\frac{dn_s }{d\ln k}
	&=	16\varepsilon_V  \eta_V -24\varepsilon_V^2-2\zeta_V^2,
	\label{slow roll results}
}
within the slow roll approximations, where $A_s$, $r$, $n_s$, and $\frac{dn_s }{d\ln k}$ are the scalar power spectrum amplitude, tensor-to-scalar power ratio, scalar spectral index, and its running.
The current observational bounds by Planck 2018 are \cite{Aghanim:2018eyx,Akrami:2018odb}
\aln{
A_s
	&=	2.101^{+0.031}_{-0.034}\times 10^{-9},&
&&
&(68\%\ {\rm CL})\nn
r	&<	0.056,&
&&
&(95\%\ {\rm CL})\nn
n_s 
	&=	0.9665\pm 0.0038,&
\frac{dn_s }{d\ln k}
	&=	0.013\pm 0.024,&
&(68\%\ {\rm CL})
	\label{values of cosmological observables}
}
at the pivot scale $k_* =0.05\,{\rm Mpc}^{-1}$.

Under the slow-roll approximation, the scalar amplitude $A_s$ is given by Eq.~\eqref{slow roll results}.
We note that $r$ and the value of potential $U$ is related each other by fixing $A_s$ to the observed value:
\aln{U_{}\sim 1.5\times 10^{-9}\left[r\over0.05\right]M_\text{P}^{4};
\label{typical potential}
}
see also Ref.~\cite{Hamada:2017sga}.

\section{Ordinary Higgs inflation without criticality}\label{ordinary Higgs inflation results}

For ${\xi h^2\over M_\text{P}^2}\gg1$ we have simple relations for the slow roll parameters  
\aln{
\varepsilon_{V}
	&\simeq
		\frac{4}{3}\exp\left(-2\sqrt{\frac{2}{3}}\frac{\chi}{M_\text{P} }\right)\simeq \frac{3}{4N^2},&
\eta_{V}
	&\simeq
		-\frac{4}{3}\exp\left(-\sqrt{\frac{2}{3}}\frac{\chi}{M_\text{P} }\right)=-\frac{1}{N}, 
\label{conventional predictions}
}
and they provide one of the best fits to the CMB observations for the reasonable values of e-folding $N=50\text{--} 60$ (corresponding to the pivot scale).

Qualitatively, the typical value of $\xi$ can be estimated as follows. 
%
Putting the potential~(\ref{Vinf}) with $\lambda_{\rm eff} \sim \lambda_c \sim 10^{-6}$ into Eq.~\eqref{typical potential}, one can easily check that $\xi$ is around $30$.  
%

\section{Expansion around saddle point}\label{Expansion around saddle point}
%
For qualitative understanding, it is also helpful to derive the expansion of $V_{} $ around $h_s $. 
We first expand as
\aln{
&V_{} =\frac{\lambda_c }{2}
h_s^4+\lambda_1 h_s^{3}(h-h_s )+\frac{\lambda_2 }{2}h_s^2(h-h_s )^2+\frac{\lambda_3 }{3!}h_s (h-h_s )^3+\dots,\label{expansion of VH}
%
}
where
\aln{
\lambda_1 =\lambda_c \delta,\quad \lambda_2 =3\lambda_c \delta,\quad \lambda_3 =32\lambda_c . 
\label{coefficients}
}
As we explain in Section \ref{Higgs inflation with radiative}, the Higgs potential in the Einstein frame also has a saddle point at $h=h_s\Omega$. 
We are interested in the parameter space $c_s\sim 1\ \Leftrightarrow\ \tilde{h}_s \gg h_s \sim {M_\text{P}\over\sqrt\xi} $, which guarantees the large field expansion of $\chi$ as a function of $h$.   
As a result, we have
\aln{&\frac{d h}{d\chi}\sim \frac{h}{\sqrt{6}M_\text{P} }\ \Leftrightarrow\ h\sim {M_\text{P}\over\sqrt\xi} \exp\left(\frac{\chi}{\sqrt{6}M_\text{P} }\right),
\\
&\varphi\sim {M_\text{P}\over\sqrt\xi} \left(1-\frac{\left({M_\text{P}\over\sqrt\xi}\right)^2}{2h^2}\right)\ \Leftrightarrow\ \varphi-h_s \sim {M_\text{P}\over\sqrt\xi} \left[1-c-\frac{1}{2}e^{-\frac{2\chi}{\sqrt{6}M_\text{P} }}\right],
}
where $\varphi:=h/\Omega$.
Then, $U_{}$ and its derivatives with respect to $\chi$ are 
\aln{
U_{}&\sim \frac{\lambda_c }{4}h_s^4=\frac{\lambda_0 c^4}{4}\left({M_\text{P}\over\sqrt\xi}\right)^4,\label{height VE}
\\
\frac{\partial U_{}}{\partial \chi}=&\frac{\partial \varphi}{\partial \chi}\frac{\partial U_{}}{\partial \varphi}
\sim \frac{{M_\text{P}\over\sqrt\xi} e^{-2\chi/(\sqrt{6}M_\text{P} )}}{\sqrt{6}M_\text{P} }\left(\lambda_1 h_s^{3}+\lambda_2 h_s^2(\varphi-h_s )+\frac{\lambda_3 h_s }{2}(\varphi-h_s )^2\right)
\nn
&\sim \frac{\lambda_3 {M_\text{P}\over\sqrt\xi}  h_s^3}{\sqrt{6}M_\text{P} }e^{-2\chi/(\sqrt{6}M_\text{P} )}\left[\frac{\delta }{32}
+\frac{3\delta }{32c}\left(1-c-\frac{1}{2}e^{-\frac{2\chi}{\sqrt{6}M_\text{P} }}\right)
+\frac{1}{2c^2}\left(1-c-\frac{1}{2}e^{-\frac{2\chi}{\sqrt{6}M_\text{P} }}\right)^2\right],
\\
\frac{\partial^2 U_{}}{\partial \chi^2}
\sim &-\frac{\lambda_3 {M_\text{P}\over\sqrt\xi} h_s^3}{3M_\text{P}^2}e^{-2\chi/(\sqrt{6}M_\text{P} )}\Bigg[
\frac{\delta }{32}
+\frac{3\delta }{32c}\left(1-c-\frac{1}{2}e^{-\frac{2\chi}{\sqrt{6}M_\text{P} }}\right)
+\frac{1}{2c^2}\left(1-c-\frac{1}{2}e^{-\frac{2\chi}{\sqrt{6}M_\text{P} }}\right)^2
\nn
&\h{4cm}-\frac{e^{-2\chi/(\sqrt{6}M_\text{P} )}}{2}\left\{\frac{3\delta }{32}+\frac{1}{c^2}\left(1-c-\frac{1}{2}e^{-\frac{2\chi}{\sqrt{6}M_\text{P} }}\right)
\right\}
\Bigg],
}
where we have used Eq.~(\ref{expansion of VH}). 
When $c\sim 1$, the slow roll parameters are approximately given by   
\aln{
\varepsilon=\frac{M_\text{P}^2}{2}\left(\frac{U_{}^{\prime}}{U_{}}\right)^2\sim \frac{4}{3}\left(\frac{\delta }{c}\right)^2e^{-4\chi/(\sqrt{6}M_\text{P} )},\quad \eta=M_\text{P}^2\frac{U_{}^{\prime\prime}}{U_{}}\sim - \frac{4\delta }{3c}e^{-2\chi/(\sqrt{6}M_\text{P} )},
\label{approximate slow roll}
}
where $a=\lambda_1 /\lambda_0 =1+(\beta_\lambda /4\lambda)|_{\phi=\phi_s } $.  
Compared to the conventional case, we have additional suppression factor $\delta$ thanks to the saddle potential.  
Note that, when $c\ll 1$, we can no longer trust Eq.~(\ref{approximate slow roll}).   
%

%
%
%
%

\bibliography{Bibliography}
\bibliographystyle{utphys}

\end{document}